\documentclass[]{pasj02} 
\usepackage[switch,mathlines]{lineno} % add line number to manuscript

\usepackage{natbib} 
\usepackage[nooneline]{subfigure}
\usepackage{threeparttable}
\usepackage{amsmath}
\usepackage{bm}
\usepackage{comment}
\usepackage{url}
\usepackage{xcolor}
\usepackage{color}
\usepackage[normalem]{ulem}

\Received{2025/11/20}%{yyyy/mm/dd}
\Accepted{2026/5/20}%{yyyy/mm/dd}
\begin{document}

\title{Evolution of compressed clouds formed by filament coalescence. I. Oblique collisions}

%%% begin:list of authors
% Do NOT capitalize all letters in "textsc".
\author{
 Raiga \textsc{Kashiwagi},\altaffilmark{1,2}\altemailmark\orcid{0000-0002-1461-3866} \email{raiga.kashiwagi@gmail.com} 
 Kazunari \textsc{Iwasaki},\altaffilmark{2,3,4}\orcid{0000-0002-2707-7548}
 Kohji \textsc{Tomisaka},\altaffilmark{2,4}\orcid{0000-0003-2726-0892}
 and 
 Tsuyoshi \textsc{Inoue}\altaffilmark{5}\orcid{0000-0002-7935-8771} 
}
\altaffiltext{1}{Korea Astronomy and Space Science Institute, Daejeon
34055, Republic of Korea}
\altaffiltext{2}{National Astronomical Observatory of Japan, 2-21-1 Osawa, Mitaka, Tokyo 181-8588, Japan}
\altaffiltext{3}{Center for Computational Astrophysics, National Astronomical Observatory of Japan, 2-21-1 Osawa, Mitaka, Tokyo 181-8588, Japan}
\altaffiltext{4}{Astronomical Science Program, The Graduate University for Advanced Studies (SOKENDAI), 2-21-1 Osawa, Mitaka, Tokyo 181-8588,
Japan}
\altaffiltext{5}{Department of Physics, Konan University, Okamoto 8-9-1, Higashinada-ku, Kobe 658-8501, Japan}
%\footnotetext[$\dag$]{Present address: ....}

%%% end:list of authors

%% !!! Select 3 to 5 words from PASJ's key words !!! 
%% List of Key Words: https://academic.oup.com/pasj/pages/Pasj_Keywords 
%% "\KeyWords{ }" always has to be placed before ``\maketitle'' 

\maketitle

\begin{abstract}
Stars are thought to form predominantly within filamentary molecular clouds. 
Recent studies have suggested that active star formation, including the formation of stellar clusters and massive stars, occurs within so-called “hub” structures, where multiple filaments converge. 
Understanding the formation and evolution of such hub–filament systems is therefore essential for unveiling the physical processes responsible for cluster and massive star formation, although the full picture remains incomplete.
To address this, we have focused on filament–filament collisions as a potential formation mechanism of the hubs. 
In this study, we investigate the fundamental evolutionary processes of oblique collisions between two magnetized filaments using three–dimensional ideal magnetohydrodynamical simulations.
As a model of initial filaments, we consider two identical finite-length magnetized filaments, varying the collision angle between their long axes, the collision velocity, which is set perpendicular to the long axes, and the initial line mass.
We find that as the collision angle decreases from orthogonal to parallel, the compressed cloud becomes more prone to gravitational collapse. 
In addition, the instability of the post-collision compressed cloud can be explained by its energy balance. 
Specifically, if the absolute value of the gravitational energy exceeds the sum of the kinetic, thermal, and magnetic energies immediately after the collision, the cloud undergoes gravitational collapse. Conversely, if the gravitational energy is smaller, the cloud expands.
In addition, we estimate the upper limit of the collision velocity that enables hub–filament formation and identify the collision conditions favorable for massive star formation.
\end{abstract}

\pagewiselinenumbers 

\section{Introduction}
The dense structures within molecular clouds are generally organized into elongated filaments.
Observations with the {\it Herschel Space Telescope} \citep{2010A&A...518L...1P} have revealed the ubiquity of such filamentary structures \citep{2010A&A...518L.102A,2011A&A...529L...6A}.
Moreover, a large fraction of protostars are found along these filaments \citep[e.g.,][]{2015A&A...584A..91K,2019A&A...621A..42A}.
From these findings, filamentary molecular clouds are widely recognized as the birthplaces of stars \citep[e.g.,][]{2014prpl.conf...27A,2023ASPC..534..153H,2023ASPC..534..233P}.
Stars tend to form in filaments with line masses exceeding the critical value, $\lambda_\mathrm{crit}=2c_s^2/G\simeq16.7\,M_\odot\,\mathrm{pc}^{-1}$, where $c_s\simeq0.19~\mathrm{km\,s^{-1}}$ and $G$ denote the isothermal sound speed and the gravitational constant, respectively \citep{1987PThPh..77..635N,1992ApJ...388..392I}.
This critical line mass, derived from the hydrostatic equilibrium of an isothermal filament \citep{1963AcA....13...30S,1964ApJ...140.1056O}, represents the maximum line mass that can be supported by thermal pressure against self-gravity and is considered as a key parameter for the onset of star formation.

Filaments are not always isolated structures, but can also form web-like networks containing junctions where multiple filaments intersect, known as "hub-filament systems" or "hubs" \citep{2009ApJ...700.1609M}.
Many observations have indicated that active massive star formation often takes place within hub-filament systems  \citep[e.g.,][]{2010A&A...520A..49S,2013A&A...555A.112P,2014A&A...561A..83P,2020A&A...642A..87K,2022A&A...658A.114K,2021MNRAS.508.2964A,2023MNRAS.522.3719L,2024MNRAS.527.4244S,2024ApJ...967..151S}.
In particular, \citet{2020A&A...642A..87K} revealed that, in high-mass star-forming regions, all luminous clumps are located within these hubs, i.e., all massive stars preferentially form in hub–filament systems.
Moreover, this characteristic is also identified in the Large Magellanic Cloud \citep{2019ApJ...886...15T, 2023ApJ...955...52T}.
Therefore, understanding the formation processes of such hub-filament systems is deeply connected to the formation of massive stars.

Hub–filament systems are observed not only in massive star-forming regions but also in low- and intermediate-mass star-forming regions \citep{2012A&A...541A..63P, 2014ApJ...791L..23N, 2020A&A...638A..74L,2023MNRAS.520.4646K}, although some regions such as Taurus \citep{2013A&A...550A..38P} show solar-mass star formation occurring within a single filament.
Accordingly, understanding the formation of hub–filament systems is of general importance for star formation across the low- to high-mass range.

In addition, observations have revealed the existence of magnetic field structures within the hub–filament systems \citep[e.g.,][]{2020ApJ...905..158W,2021A&A...647A..78A, 2025ApJ...986...48J,2025A&A...703A..74K}. 
Magnetic fields not only act as a force against gravity but also influence gas dynamics by guiding the motion of gas along magnetic field lines. 
Therefore, magnetic fields are expected to play a crucial role in the formation and evolution of hub–filament systems.

One proposed formation mechanism for hub structures is the coalescence of multiple filaments \citep[e.g.,][]{2019ApJ...886...15T, 2020A&A...642A..87K,2022A&A...658A.114K}.
Similarly, for cluster formation, filament collisions have been proposed as a promising mechanism \citep[e.g.,][]{2011A&A...528A..50D,2014ApJ...791L..23N}.
Observational studies, particularly those based on gas kinematics, have suggested the existence of filament collisions \citep[e.g.,][]{2014ApJ...797...58D,2015A&A...574L...6F,2017ApJ...851..140D,2017ApJ...835..108S,2019A&A...631A...3M,2019ApJ...884...84D,2022MNRAS.513.2942D,2023ApJ...958...17P,2024MNRAS.527.5895D,2025AJ....169...56M,2025A&A...701A.248S,2025ApJ...994..118M}.

Numerical simulations of filament–filament collisions have been conducted in recent years \citep{2011A&A...528A..50D,2021MNRAS.507.3486H,2023ApJ...954..129K,2024MNRAS.532L..42H,2024ApJ...974..265K}. 
In particular, the fundamental processes of filament collisions have been explored by \citet{2023ApJ...954..129K} (hereafter RK23), which examined parallel (head-on) collisions, and \citet{2024ApJ...974..265K} (hereafter RK24), which focused on orthogonal collisions.
These studies investigated the evolution of the shocked clouds formed by filament collisions and identified the conditions under which these clouds undergo gravitational collapse.

RK23 investigated head-on collisions between two identical, infinitely long filaments that were in magnetohydrostatic equilibrium and threaded by a lateral magnetic field. 
The models explored a range of relative velocities from the sound speed up to ten times that value. In this configuration, the shocked cloud evolved into a filament-like structure. 
The study found that when the total line mass exceeds the magnetic critical line mass \citep{2014ApJ...785...24T,2021ApJ...911..106K}, the shocked cloud undergoes gravitational collapse, whereas cases with a total line mass lower than the magnetic critical value remain stable.

RK24 extended this analysis to orthogonal collisions, adopting the same filament model but setting the angle between the axes to $\pi/2$. 
In this case, the shocked cloud became ellipsoidal rather than filamentary. 
The stability of this cloud was found to depend on both the collision velocity and the magnetic support: for low-velocity collisions, the collapse condition follows the magnetic critical mass of a sphere \citep{1976ApJ...210..326M,1988ApJ...335..239T}, whereas for higher velocities, the evolution of the shocked region is determined by the initial velocity.
From an energy balance analysis performed immediately after the collision, RK24 showed that the cloud undergoes collapse only when the magnitude of its gravitational energy exceeds the sum of its kinetic, thermal, and magnetic energies; otherwise, it expands.

However, the previous studies, RK23 and RK24, were restricted to limited configurations—parallel and orthogonal collisions, respectively.
In this study, we extend these works by systematically varying the angle between the filament axes to explore oblique collisions.
This approach allows us to provide a more comprehensive understanding of the fundamental processes involved in filament collisions.

We examine collisions between less-massive filaments, aiming to form relatively low-mass hubs and to clarify how the collision angle affects the outcomes. 
Based on these results, we then extend the discussion to collisions leading to the formation of more massive hubs.

The structure of this paper is as follows.
Section \ref{sec:model} introduces the numerical models and parameters used in our simulations, where we adopt relatively low-mass filaments as the initial conditions.
Section \ref{sec:method} describes the numerical methods used in our simulations.
Section \ref{sec:result} explains the results of oblique collision simulations.
In section \ref{sec:discussion}, we first examine the stability of the shocked cloud, then derive the formation conditions for low-mass hub-filament systems, and finally discuss the implications for the formation of massive hub-filament systems.
Finally, section \ref{sec:summary} provides a summary and the conclusions of this study.

\section{Models and parameters}\label{sec:model}
This section outlines our models for investigating oblique collisions of two finite, identical filaments penetrated by a lateral magnetic field. 

Following RK24, as an initial filament model, we adopt the magnetohydrostatic equilibrium filament derived by \citet{2014ApJ...785...24T}, a self-consistent magnetohydrostatic solution for a filament threaded by a magnetic field perpendicular to its axis.
In this section, we primarily describe the modifications implemented in this study.

\begin{figure}
    \includegraphics[keepaspectratio,scale=0.4]{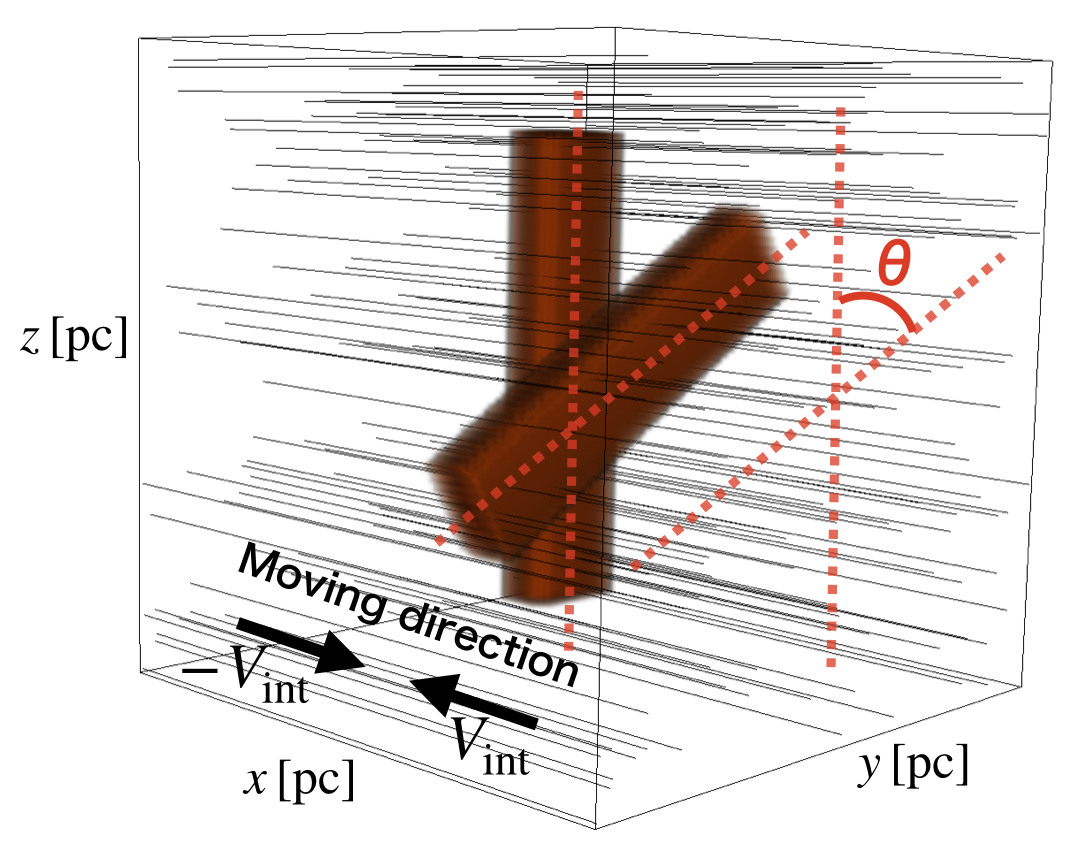}
      \caption{
      An example of the initial condition for oblique collisions.
      The simulation domain is a cubic box with a side length of $L_\mathrm{box} = 2.2~\mathrm{pc}$ (i.e., $x = y = z = 2.2~\mathrm{pc}$).
      Each filament is initially in magnetohydrostatic equilibrium. 
      The angle between the major axes of the filaments is denoted by $\theta$.
      Magnetic field lines (black lines) are globally aligned along the x-axis, and the initial speed $\left(V_\mathrm{int}\right)$ is also given in the x-direction.
      {Alt text: A schematic 3D view of the simulation’s initial condition for an oblique collision between two magnetized filaments inside a cubic box with 2.2 pc sides.}}
    \label{fig:initial_condition}
\end{figure} 

In this study, we adopt finite-length filaments because considering infinitely long filaments, as in RK23 and RK24, makes the treatment of boundary conditions difficult when the filaments collide with an arbitrary angle. 
The finite-length configuration enables a more manageable numerical setup.

Figure \ref{fig:initial_condition} illustrates one of the initial setups for the oblique collisions, shown in the Cartesian coordinate system $\left(x,\,y,\,z\right)$.
The two identical finite filaments are placed adjacent to each other, sharing a common magnetic flux tube.
One filament is inclined by an angle $\theta$ relative to the other, and they collide along the $\pm x$ direction.
Thus, when $\theta = 90^\circ$, the long axis of the $x>0$ filament is along the $y$-axis (orthogonal), and when $\theta =0^\circ$, it becomes aligned with the $z$-axis (parallel).

\begin{longtable}{ccccccc}
 \caption{Model Parameters and Results for the Oblique Collisions. The plasma beta is set to $\beta_0 = 0.1$ (i.e., $B_0=19\,\mu \mathrm{G}$) in all models.}\label{tab:parameters}  
\hline\noalign{\vskip3pt} 
  Model\footnotemark[(1)] &  $\theta$\footnotemark[(2)] & $\lambda_0/\lambda_\mathrm{crit,B}$\footnotemark[(3)]& $V_\mathrm{int}/c_s$\footnotemark[(4)] & $M_\mathrm{hub}\,[M_\odot]$\footnotemark[(5)] & $t_e \,\mathrm{[Myr]}$\footnotemark[(6)] & Result\footnotemark[(7)]\\   [2pt] 
\hline\noalign{\vskip3pt} 
\endfirsthead      
\hline\noalign{\vskip3pt} 
  Model\footnotemark[(1)] &  $\theta$\footnotemark[(2)] & $\lambda_0/\lambda_\mathrm{crit,B}$\footnotemark[(3)]& $V_\mathrm{int}/c_s$\footnotemark[(4)] & $\rho_c/\rho_s$\footnotemark[(5)] & $t_e \,\mathrm{[Myr]}$\footnotemark[(6)] &Result\footnotemark[(7)]\\  [2pt] 
\hline\noalign{\vskip3pt} 
\endhead
\hline\noalign{\vskip3pt} 
\endfoot
\hline\noalign{\vskip3pt} 
\multicolumn{2}{@{}l@{}}{\hbox to0pt{\parbox{160mm}{\footnotesize
\hangindent6pt\noindent
\hbox to6pt{\footnotemark[(1)]\hss}\unskip% 
  Name of the model. \\
\hangindent6pt\noindent
%\hbox to6pt{\footnotemark[(2)]\hss}\unskip% 
%  Plasma beta value.  \\
%\hangindent6pt\noindent
\hbox to6pt{\footnotemark[(2)]\hss}\unskip% 
  Collision angle.  \\
\hangindent6pt\noindent
\hbox to6pt{\footnotemark[(3)]\hss}\unskip% 
  Initial line mass normalized by the magnetic critical line mass.  \\
\hangindent6pt\noindent
\hbox to6pt{\footnotemark[(4)]\hss}\unskip% 
  Collision velocity.  \\
\hangindent6pt\noindent
\hbox to6pt{\footnotemark[(5)]\hss}\unskip% 
  Mass of the shocked cloud estimated by equation (\ref{eq:Delta_M_vs_theta}).  \\
\hangindent6pt\noindent
\hbox to6pt{\footnotemark[(6)]\hss}\unskip% 
 The time when the numerical density limit is reached. Models marked with “$>$” reached $t_{\mathrm{stop}}$ before reaching $n_{\mathrm{lim}}$.  \\
\hangindent6pt\noindent
\hbox to6pt{\footnotemark[(7)]\hss}\unskip% 
  Evolutionary mode of the shocked cloud. 
}\hss}} 
\endlastfoot 
  %B1L04V02th00 & 1.0 & $0$ & 0.4 & $\pm1.0$ & ccccc & Collapse \\
  %B1L04V18th00 & 1.0 & $0$ & 0.4 & $\pm9.0$ & ccccc & Expansion \\
  %B1L05V02th30 & 1.0 & $\pi/6$ & 0.5 & $\pm1.0$ & ccccc & Collapse \\
  %B1L05V10th30 & 1.0 & $\pi/6$ & 0.5 & $\pm5.0$ & ccccc & Expansion \\
  %B1L05V02th60 & 1.0 & $\pi/3$ & 0.5 & $\pm1.0$ & ccccc & Collapse \\
  %B1L05V10th60 & 1.0 & $\pi/3$ & 0.5 & $\pm5.0$ & ccccc & Expansion \\
  %B1L05V02th90 & 1.0 & $\pi/2$ & 0.5 & $\pm1.0$ & ccccc & Collapse \\
  %B1L05V10th90 & 1.0 & $\pi/2$ & 0.5 & $\pm5.0$ & ccccc & Expansion \\
  L04V01th30 & $\pi/6$ & $0.4$ & $0.5$ & $21.8$ & $1.15$ & Collapse  \\
  L04V02th30 & $\pi/6$ & $0.4$ & $1.0$ & $21.8$ & $1.24$ & Collapse  \\
  L04V06th30 & $\pi/6$ & $0.4$ & $3.0$ & $21.8$ & $1.80$ &Collapse  \\
  L04V10th30 & $\pi/6$ & $0.4$ & $5.0$ & $21.8$ & $>2.18$ &Expansion \\
  L04V14th30 & $\pi/6$ & $0.4$ & $7.0$ & $21.8$ & $>2.18$ &Expansion \\
  L05V01th30 & $\pi/6$ & $0.5$ & $0.5$ & $27.3$ & $0.58$ &Collapse  \\
  L05V02th30 & $\pi/6$ & $0.5$ & $1.0$ & $27.3$ & $0.57$ &Collapse  \\
  L05V10th30 & $\pi/6$ & $0.5$ & $6.5$ & $27.3$ & $0.76$ &Collapse  \\
  L05V13th30 & $\pi/6$ & $0.5$ & $6.5$ & $27.3$ & $1.32$ &Expansion \\
  L05V17th30 & $\pi/6$ & $0.5$ & $8.5$ & $27.3$ & $>2.18$ &Expansion \\
  L04V01th60 & $\pi/3$ & $0.4$ & $0.5$ & $12.6$ & $0.97$ &Collapse  \\
  L04V02th60 & $\pi/3$ & $0.4$ & $1.0$ & $12.6$ & $0.95$ &Collapse  \\
  L04V06th60 & $\pi/3$ & $0.4$ & $3.0$ & $12.6$ & $>2.18$ &Expansion \\
  L04V10th60 & $\pi/3$ & $0.4$ & $5.0$ & $12.6$ & $>2.18$ &Expansion \\
  L05V01th60 & $\pi/3$ & $0.5$ & $0.5$ & $15.7$ & $0.52$ &Collapse  \\
  L05V02th60 & $\pi/3$ & $0.5$ & $1.0$ & $15.7$ & $0.52$ &Collapse  \\
  L05V08th60 & $\pi/3$ & $0.5$ & $4.0$ & $15.7$ & $0.57$ &Collapse  \\
  L05V13th60 & $\pi/3$ & $0.5$ & $6.5$ & $15.7$ & $1.60$ &Expansion \\
  L05V16th60 & $\pi/3$ & $0.5$ & $8.0$ & $15.7$ & $>2.18$ &Expansion \\
  L04V01th90 & $\pi/2$ & $0.4$ & $0.5$ & $10.9$ & $0.93$ &Collapse  \\
  L04V02th90 & $\pi/2$ & $0.4$ & $1.0$ & $10.9$ & $0.86$ &Collapse  \\
  L04V06th90 & $\pi/2$ & $0.4$ & $3.0$ & $10.9$ & $>2.18$ &Expansion \\
  L04V10th90 & $\pi/2$ & $0.4$ & $5.0$ & $10.9$ & $>2.18$ &Expansion \\
  L05V01th90 & $\pi/2$ & $0.5$ & $0.5$ & $13.6$ & $0.50$ &Collapse  \\
  L05V02th90 & $\pi/2$ & $0.5$ & $1.0$ & $13.6$ & $0.50$ &Collapse  \\
  L05V08th90 & $\pi/2$ & $0.5$ & $4.0$ & $13.6$ & $0.55$ &Collapse  \\
  L05V13th90 & $\pi/2$ & $0.5$ & $6.5$ & $13.6$ & $1.92$ &Expansion \\
  L05V18th90 & $\pi/2$ & $0.5$ & $9.0$ & $13.6$ & $>2.18$ &Expansion \\ 
\end{longtable}

To simulate oblique collisions, we must provide the density structure and magnetic field distribution for arbitrary collision angles $\theta$ between the long axes of the two filaments. 
In this study, the filament in the $x<0$ region is placed with its long axis aligned with the $z$-axis, and therefore the initial filament structure is obtained by stacking the two-dimensional profile of the magnetohydrostatic equilibrium filament along the $z$-axis.
In contrast, the other filament in the $x>0$ region is obtained by rotating the structure of the filament in the $x<0$ region by $\theta$ about the $x$-axis.

The initial magnetic field $\bm{B}$ is obtained by taking the curl of a vector potential $\bm{A}$
(i.e., $\bm{B}=\nabla\times \bm{A}$), so that $\bm{\nabla}\cdot \mathbf{B}$ remains zero in round-off errors. 
To achieve this, the vector potential of the magnetohydrostatic equilibrium solution is divided into two components: a uniform magnetic field component ($\bm{A}_0$) and a residual component ($\bm{A}_\mathrm{curv}$). 
In this model, since the uniform magnetic field component is along the $x$-axis, the corresponding vector potential is expressed as $\bm{A}_0=(0,0,B_0y)$, where $B_0$ denotes the strength of the uniform component of $\bm{B}$.
For the filament with $x < 0$, the residual vector potential is expressed as $\bm{A}_\mathrm{curv} = (0,0,A^{\mathrm{3D,\,curve}}_z(x,y))$, where $A^{\mathrm{3D,\,curve}}_z(x,y)$ represents the bending component of the magnetic field. Here, $A^{\mathrm{3D,\,curve}}_z(x,y)$ is obtained by replicating the two-dimensional vector potential $A^{\mathrm{2D,\,curve}}_z(x,y) = A^{\mathrm{2D}}_z(x,y) - B_0 y$ along the $z$-axis, where $A^{\mathrm{2D}}_z(x,y)$ corresponds to the vector potential of a magnetohydrostatic equilibrium filament.
For the filament with $x > 0$, the residual component is expressed as $\bm{A}_\mathrm{curv}=(0,-A^{\mathrm{3D,\,curve}}_{z}(x,y)\sin\theta,\,A^{\mathrm{3D,\,curve}}_{z}(x,y)\cos\theta)$, where this form is obtained by rotating the expressions for  $\bm{A}_\mathrm{curv}=(0,0,\,A^{\mathrm{3D,\,curve}}_z(x,y))$ at $x < 0$ by $\theta$ around the $x$-axis. 
After deriving the curved component of the magnetic field, $\bm{B}_\mathrm{curv}$, from the vector potential, the initial magnetic field $\bm{B}$ is obtained by adding $\bm{B}_\mathrm{curv}$ to the uniform magnetic field $\bm{B}_0 =\nabla\times\bm{A}_0= (B_0, 0, 0)$.

In our configurations, filament collisions are assumed to occur along the magnetic field lines.
More generally, collisions that compress the magnetic field are also possible. 
For example, RK23 also examined two-dimensional head-on collisions of filaments that do not share the same magnetic flux. 
In their models, compared to the case of shared magnetic flux, twice as much magnetic flux is contained in the shocked region, allowing a larger amount of gas to be sustained against gravity.
Therefore, the magnetic field geometry adopted here represents a configuration in which magnetic effects are relatively less pronounced, and gravitational instability is more easily triggered.
%A quantitative investigation of these effects requires simulations with different magnetic field geometries and non-ideal MHD effects, which will be addressed in future work.

In addition, following RK23 and RK24, we assume that the filament is immersed in an ambient pressure and introduce a scalar field to reproduce the magnetohydrostatic equilibrium state described by \citet{2014ApJ...785...24T}.
Details of the scalar field implementation are described in RK23 and RK24.

In this study, the gas temperature of the filament is assumed to be $10\,\mathrm{K}$, corresponding to $c_s = 0.19\,\mathrm{km\,s^{-1}}$.
The initial density at the filament surface is set to $\rho_{s} = \mu m_p\,n_s=4.01 \times 10^{-21}\,\mathrm{g\,cm^{-3}}$, corresponding to a number density of  $n_s=1.0\times 10^3\,\mathrm{cm^{-3}}$, where the mean molecular weight is assumed to be 
$\mu = 2.4$ and $m_p$ denotes the proton mass.
Under these conditions, the free-fall timescale is $t_\mathrm{ff} \equiv \left(4\pi G\rho_s\right)^{-1/2} \simeq 0.55\,\mathrm{Myr}$, and the characteristic scale length is $L=c_s t_\mathrm{ff}\simeq 0.11\,\mathrm{pc}$.

In our simulations, we vary three key parameters: the collision angle $\theta$ between the filament axes, the initial line mass $\lambda_0$, and the initial speed $V_\mathrm{int}$.
We consider three collision angles: $\theta=\pi/2$  (perpendicular), $\pi/3$ (intermediately oblique), and $\pi/6$  (moderately oblique)\footnote{
For more parallel configurations (i.e., collision angles below $\pi/6$), the finite filament length in our setup prevents an ideal reproduction of the shocked region, as the compressed cloud comes into contact with the filament edges and its evolution may consequently differ. 
Since the aim of this study is to understand the basic physics of the oblique collision process under idealized conditions, we limit our analysis to cases with $\theta \ge \pi/6$, while a detailed investigation of smaller angles (e.g., $\theta = \pi/12$ or $\theta = 0$) will be presented in a subsequent paper.}. 
The plasma beta is defined as the ratio of the gas pressure to the magnetic pressure.
In this study, we specifically define the magnetic field strength $B_0$ far from the filament as $B_0 = (8\pi p_\mathrm{ext}/\beta_0)^{1/2}$, where $p_\mathrm{ext}=c_s^2\rho_s$ and $\beta_0$ denote the ambient pressure and the plasma beta of the ambient gas, respectively.
In this study, we fix $\beta_0 = 0.1$ (i.e., $B_0=19 \,\mu \mathrm{G}$) as the fiducial value in order to see the effect of the collision angle, in contrast to RK24, which considered multiple values of $\beta_0$. 
The initial line mass $\lambda_0$ of each filament is normalized by the magnetic critical line mass, 
\begin{equation}\label{eq:critical_line_mass}
\lambda_{\mathrm{crit,B}}\simeq 0.24\frac{\Phi}{G^{1/2}}+1.66\frac{c^2_s}{G},
\end{equation}
where $\Phi$ is the magnetic flux per unit length \citep{2014ApJ...785...24T}.
The magnetic flux is given by $\Phi =R_0B_0$, where $R_0$ is the radius of the parental filament. 
We fix $R_0 = 0.22\,\mathrm{pc}$, thereby minimizing the contribution of $R_0$ to $\Phi$ (see also RK23 and RK24). 
The resulting value of the magnetic critical line mass $\lambda_{\mathrm{crit,B}}$ for $\beta_0 = 0.1$ is
$\lambda_{\mathrm{crit,B}}(\beta_0 = 0.1) = 1.91\,\lambda_\mathrm{crit}$,
using the thermal critical line mass.
Accordingly, we consider two types of initial line mass, $\lambda_0$, corresponding to $0.4$ and $0.5$ times $\lambda_{\mathrm{crit,B}}$, i.e., $12.8$ and $15.9\,M_\odot\,\mathrm{pc^{-1}}$.
The initial speed ($V_\mathrm{int}$) of each filament is set from $0.5c_s$ to $9.0c_s$ (i.e., the initial relative velocity ranges from $c_s$ to $18.0c_s$), which comes from the internal turbulence of molecular clouds.

In table \ref{tab:parameters}, we summarize the model parameters used in the oblique collision simulations for a total of 28 models.
The model names are determined according to the following quantities: "$\rm L$" denotes the line mass normalized by $\lambda_\mathrm{crit,B}$, "$\rm V$" represents the initial relative speed normalized by $c_s$, and "$\rm th$" indicates the collision angle $\theta$ in degrees.
For instance, Model L04V02th30 represents filaments with $\lambda_0 = 0.4 \lambda_\mathrm{crit,B}$ and $\theta = \pi/6$, colliding with the initial speed of $V_\mathrm{int} = c_s$ for each filament.
In this table, we also list the expected hub mass in the shocked region for the present configurations. 
The hub mass is theoretically given by
\begin{equation}\label{eq:Delta_M_vs_theta}
 M_\mathrm{hub} = {4 R_0 \lambda_0}/{\sin\theta},
\end{equation}
which is derived from the geometric relationship between the collision angle and the cross-sectional area of the shocked region.
For a more general set of filament collision configurations and the resulting hub mass, see \citet{2026PASJ..tmp...53T}.

The filament length is fixed at $L_\mathrm{fil} = 1.65\,\mathrm{pc}$, which covers the typical lengths of filaments observed in nearby molecular clouds and infrared dark clouds \citep[e.g.,][]{2019A&A...621A..42A,2023ASPC..534..153H}.

For simplicity, the density profile of each filament is truncated sharply at its longitudinal ends.
Such a configuration is expected to induce edge-dominant collapse, in which the filament ends collapse preferentially \citep[e.g.,][]{1983A&A...119..109B}.
In addition, a collision between finite-length filaments has been investigated in the previous hydrodynamic simulation study \citep{2021MNRAS.507.3486H}. 
\citet{2021MNRAS.507.3486H} demonstrated that, depending on the initial relative velocity, line mass, and separation of the filaments, several evolutionary pathways are possible, including symmetric mergers without cores at the filament ends, mergers dominated by edge collapse, or complete filament collapse prior to merging. 
In particular, they found that symmetric filament mergers are favored for high relative collision velocities ($>0.3\, \mathrm{km\,s^{-1}}$) and small initial separations ($<0.4\,\mathrm{pc}$), where edge effects become less significant.
Based on these results, we adopt initial conditions in which the filaments are placed in close proximity and are given relative velocities ranging from the sound speed to eighteen times the sound speed, so as to minimize the influence of edge collapse.
Consequently, the impact of the finite filament length on our results is expected to be minimal, and our analysis primarily reflects the evolution of the shocked region, especially the early phase leading to the onset of collapse, which is the main focus of this study.

\section{Numerical methods}\label{sec:method}
To solve the ideal magnetohydrodynamic (MHD) equations with self-gravity, we use Athena++ \citep{2020ApJS..249....4S} with the following numerical setup.
Time integration is performed using a fourth-order Runge–Kutta scheme \citep{2010JCoPh.229.1763K} to solve the MHD equations.
Numerical fluxes are computed using the low-dissipation Harten–Lax–van Leer discontinuities (LHLLD) method \citep{2021JCoPh.44610639M}.
Spatial reconstruction is based on the piecewise linear method (PLM) applied to the characteristic variables.
To ensure the divergence-free condition of the magnetic field, Athena++ adopts the constrained transport method \citep{1988ApJ...332..659E,2008JCoPh.227.4123G}.
The multigrid algorithm, implemented in Athena++ by \citet{2023ApJS..266....7T}, is employed to solve Poisson’s equation for the gravitational potential.
For boundary conditions, we apply outflow conditions for all MHD variables in every direction, while the gravitational potential is treated with multipole boundary conditions.

We employ the adaptive mesh refinement (AMR) technique \citep{1989JCoPh..82...64B%,1984JCoPh..53..484B
}, which is implemented 
in Athena++ by \citet{2020ApJS..249....4S}, to address issues arising from insufficient resolution during the evolution and proceeding with the collapse.
The simulation domain is a cubic box with $L_\mathrm{box}=2.2\,\mathrm{pc}$, divided into a root grid of $512^3$ cells.
The Jeans criterion \citep{1997ApJ...489L.179T} is used for a refinement condition: the grid spacing $\Delta x$ must satisfy the condition $8\Delta x \le \lambda_J$, where $\lambda_J = \sqrt{\pi c_s^2/(G\rho)}$ is the Jeans length.
The mesh refinement condition is applied up to four levels, including the root level and three finer levels.
Therefore, the finest cell size is $\simeq 0.0005\,\mathrm{pc}$ for the effective grid number of $4096^3$, and the maximum density allowed by the resolution corresponds to $n_\mathrm{lim}\simeq 2.59\times 10^7\,\mathrm{cm^{-3}}$.
The elapsed time to reach $n_\mathrm{lim}$ is defined as $t_e$. 
Even when the maximum density does not reach $n_\mathrm{lim}$, we continue the simulations up to four times $t_\mathrm{ff}$, which corresponds to $t_\mathrm{stop}=4t_\mathrm{ff}\simeq 2.18\,\mathrm{Myr}$.

We performed a resolution convergence test by varying the refinement criterion, defined as the number of cells per the local Jeans length, $n_j \equiv \lambda_J/\Delta x = 4, 8, 16$, for three representative models (L05V01th60, L05V13th60, and L05V16th60).
For model L05V01th60, which exhibits monotonic contraction, the evolution is almost the same for all resolutions $n_j\ge4$ and the elapsed time to reach $n_\mathrm{lim}$ is $t_e = 0.51\,\mathrm{Myr}$ ($n_j = 16$), $0.52\,\mathrm{Myr}$ ($n_j = 8$), and $0.51\,\mathrm{Myr}$ ($n_j = 4$).
However, in model L05V13th60, $t_e$ shows a slight dependence on $n_j$, with $t_e = 1.52\,\mathrm{Myr}$ ($n_j = 16$), $1.60\,\mathrm{Myr}$ ($n_j = 8$), and $2.30\,\mathrm{Myr}$ ($n_j = 4$), although the qualitative evolution is unchanged: the shocked cloud expands, oscillates, and then collapses.
For model L05V16th60, no collapse occurs before $t_\mathrm{stop}$ for any $n_j$.
Setting the refinement condition to $n_j \le 4$ may affect the elapsed time in some models \citep{1997ApJ...489L.179T}. 
To ensure numerical reliability, we therefore adopt $n_j = 8$ in this study. 
This setting allows us to sufficiently capture the quantitative evolution.

\begin{figure}
    \includegraphics[keepaspectratio,scale=0.4]{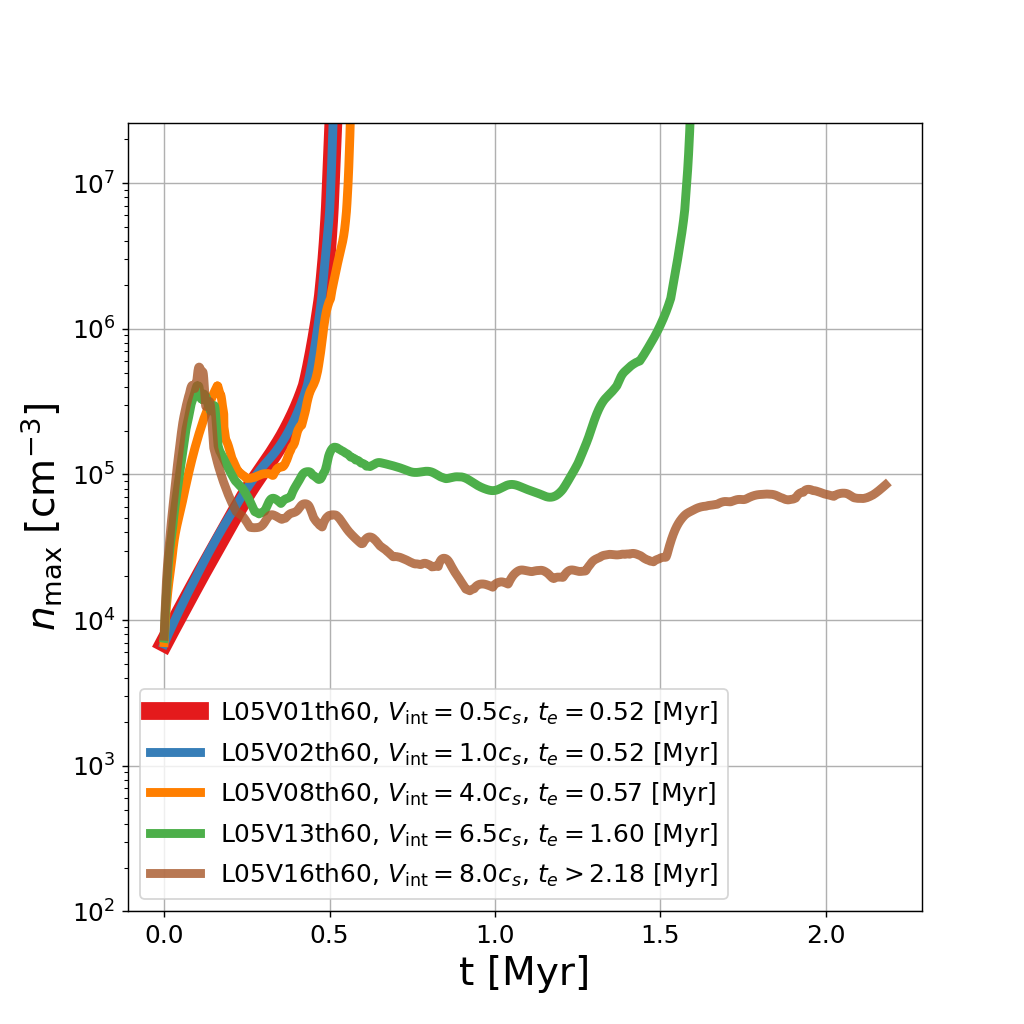}
      \caption{
      Time evolution of the maximum density. 
      The vertical axis represents the maximum density, and the horizontal axis shows time.
      The model parameters are $\lambda_0 = 0.5\lambda_{\mathrm{crit,B}}$, and $\theta = \pi/3$.
      The thick red line corresponds to the initial speed $V_\mathrm{int} = 0.5 c_s$, the blue line to $V_\mathrm{int} = c_s$, the orange line to $V_\mathrm{int} = 4.0c_s$, the green line to $V_\mathrm{int} = 6.5c_s$, and the brown line to $V_\mathrm{int} = 8.0c_s$.
      In addition, $t_e$ corresponds to the time when the density reaches the numerical limit ($n_\mathrm{lim}$).
      {Alt text: A line plot showing the time evolution of the maximum density.}
      }
    \label{fig:Maxden_vs_time_B01L05th60}
\end{figure} 
\section{Results}\label{sec:result}

\begin{figure*}[t]
%\centering
\begin{tabular}{ccc}
\subfigure[]{
\includegraphics[width=0.31\textwidth]{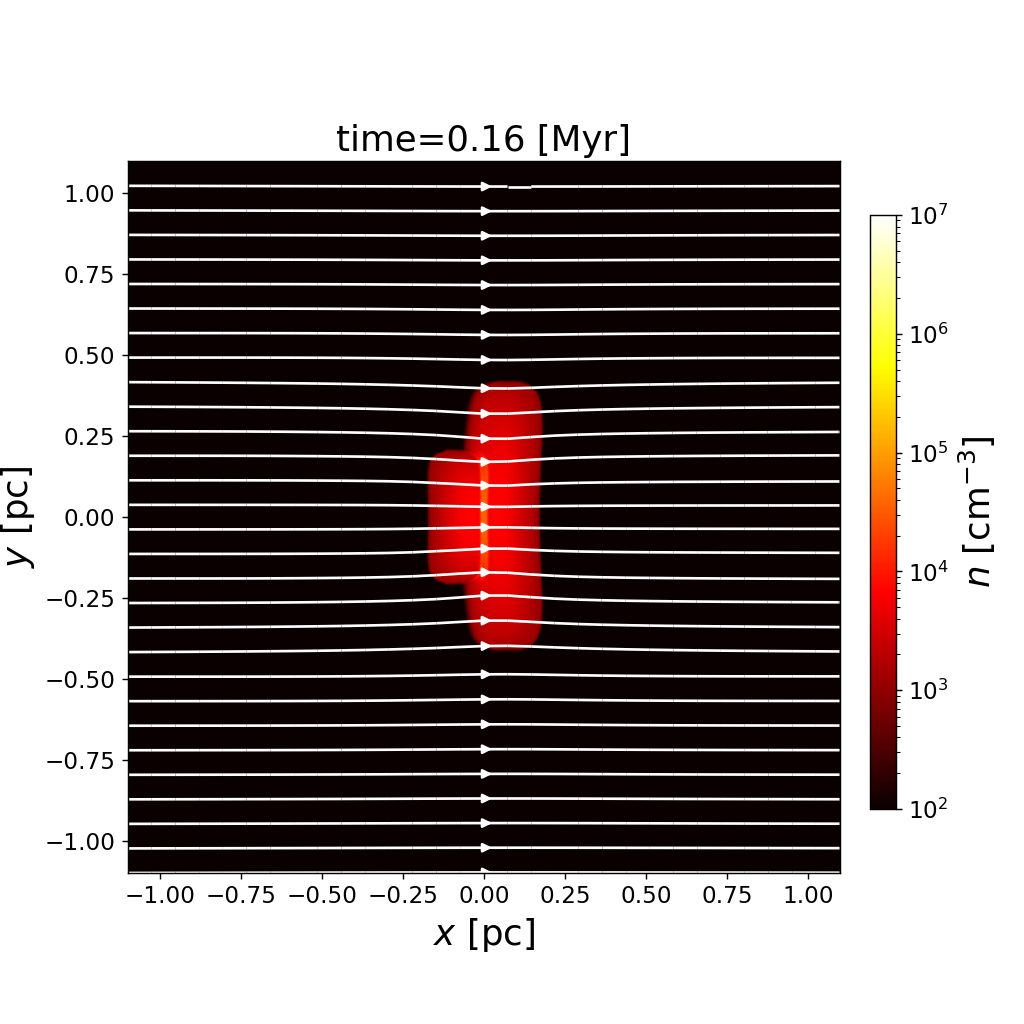}
} &
\subfigure[]{
\includegraphics[width=0.31\textwidth]{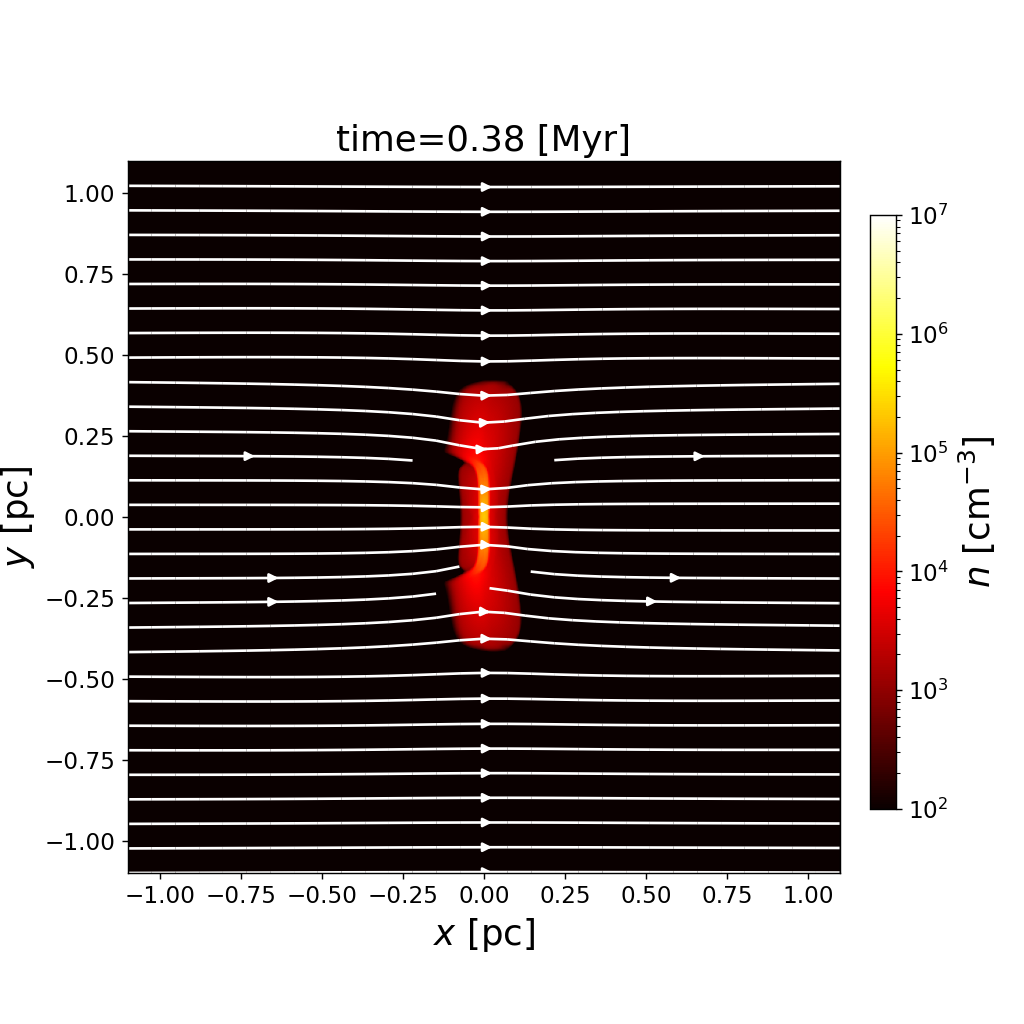}
} &
\subfigure[]{
\includegraphics[width=0.31\textwidth]{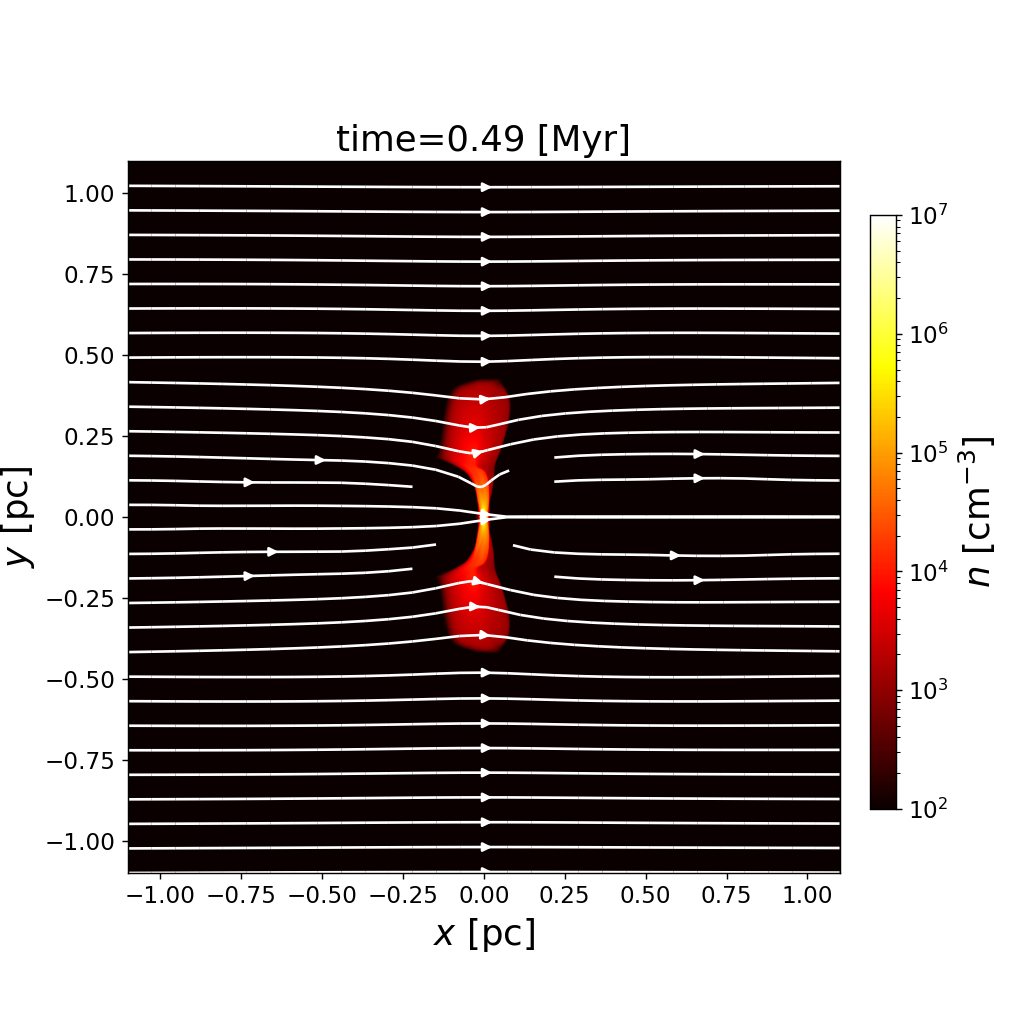}
} \\
\subfigure[]{
\includegraphics[width=0.31\textwidth]{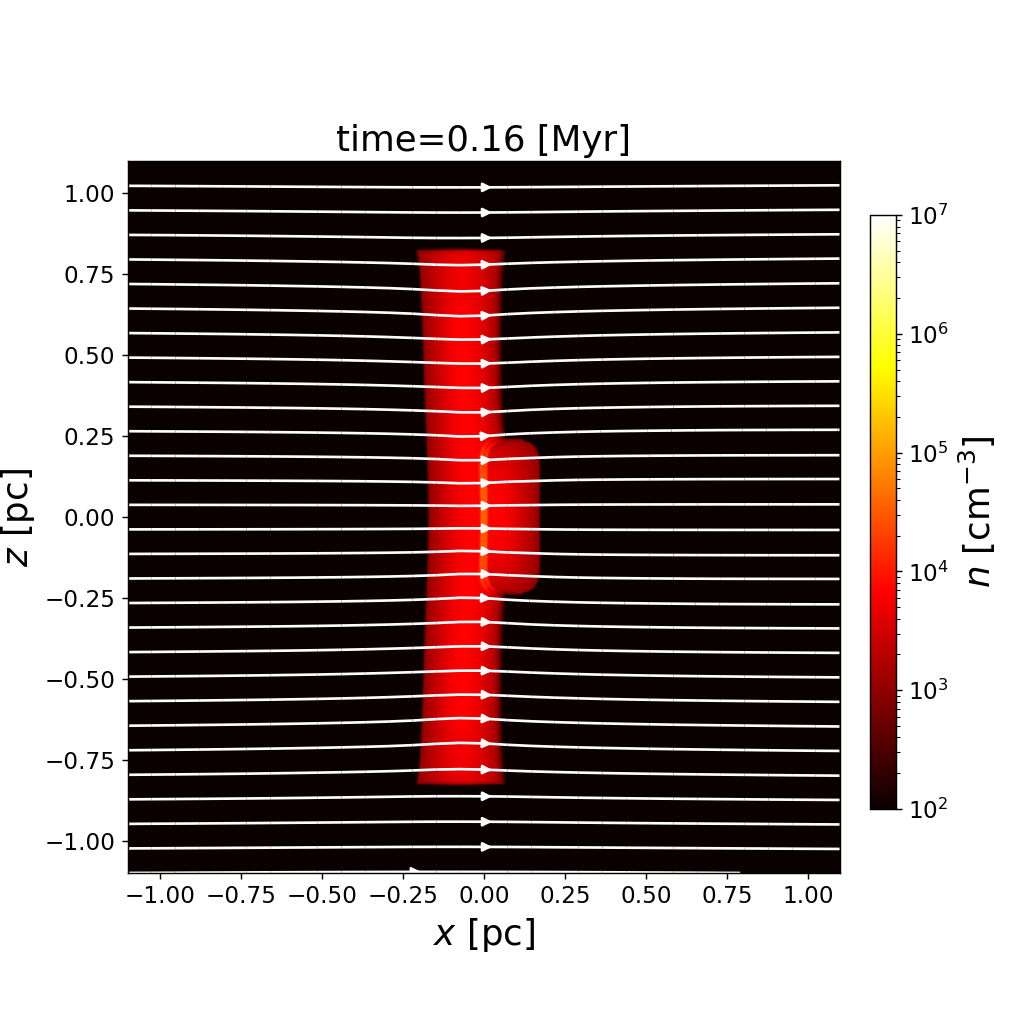}
} &
\subfigure[]{
\includegraphics[width=0.31\textwidth]{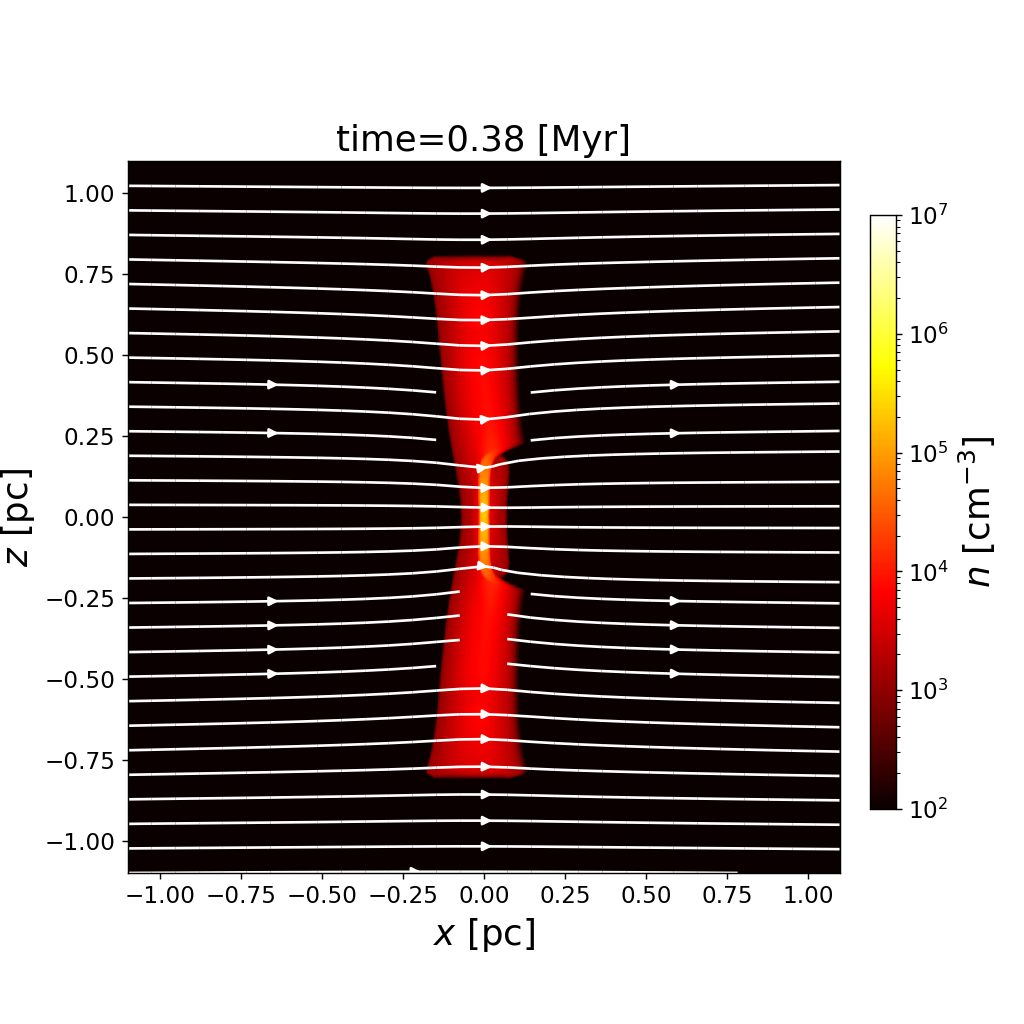}
} &
\subfigure[]{
\includegraphics[width=0.31\textwidth]{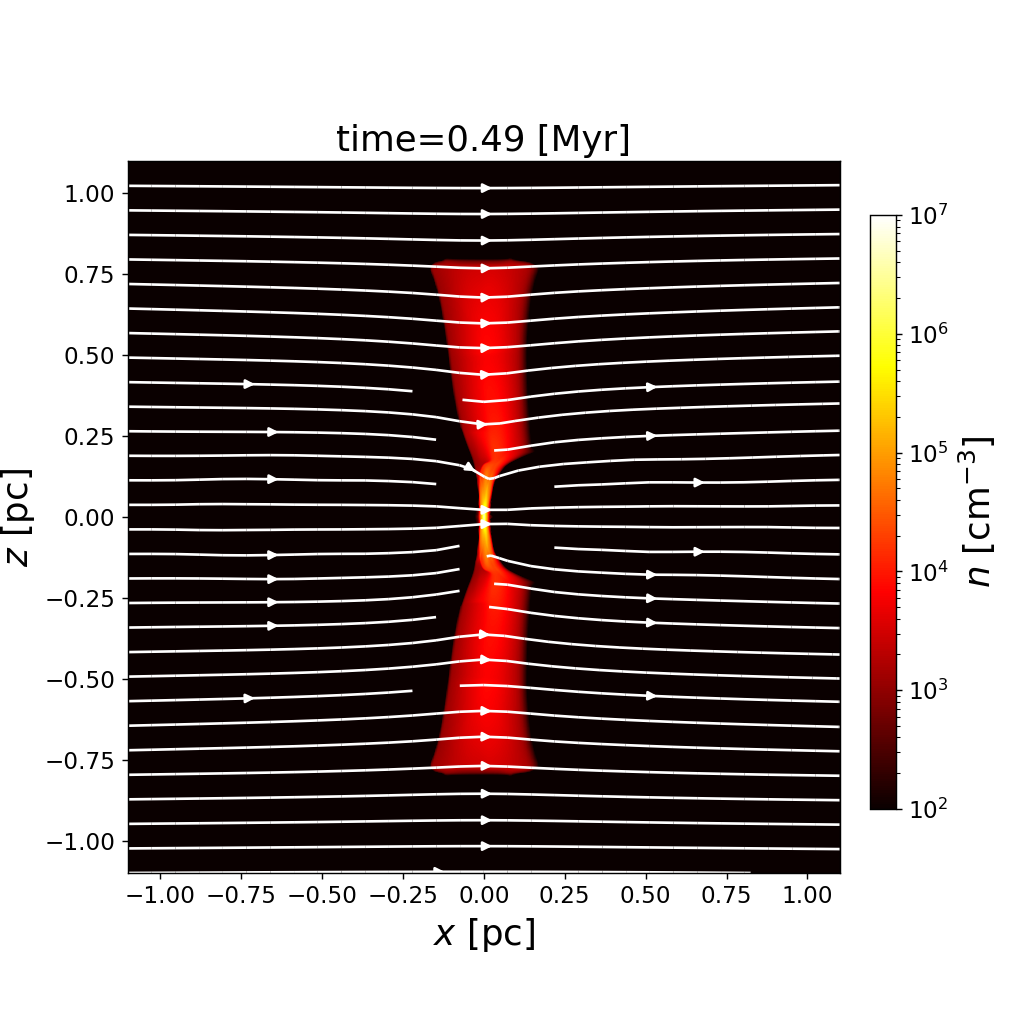}
} \\
\subfigure[]{
\includegraphics[width=0.31\textwidth]{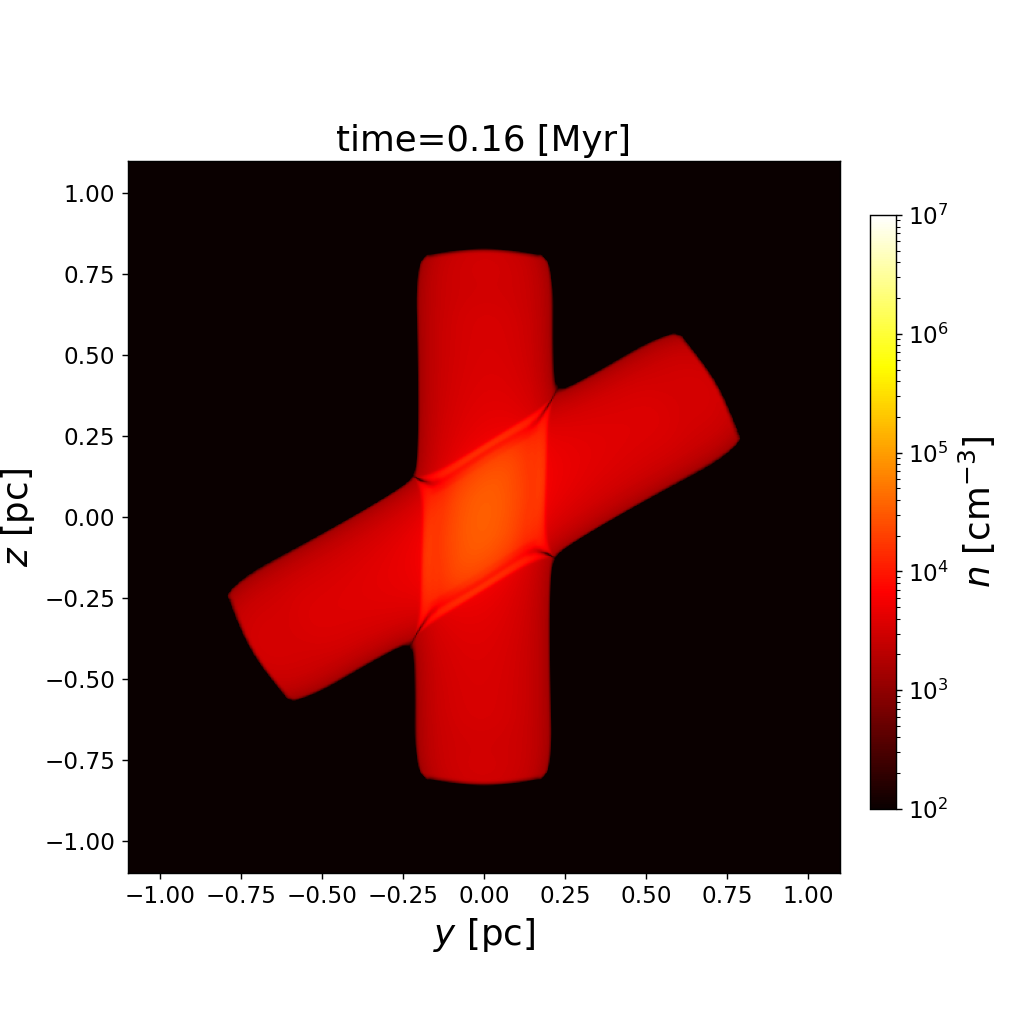}
} &
\subfigure[]{
\includegraphics[width=0.31\textwidth]{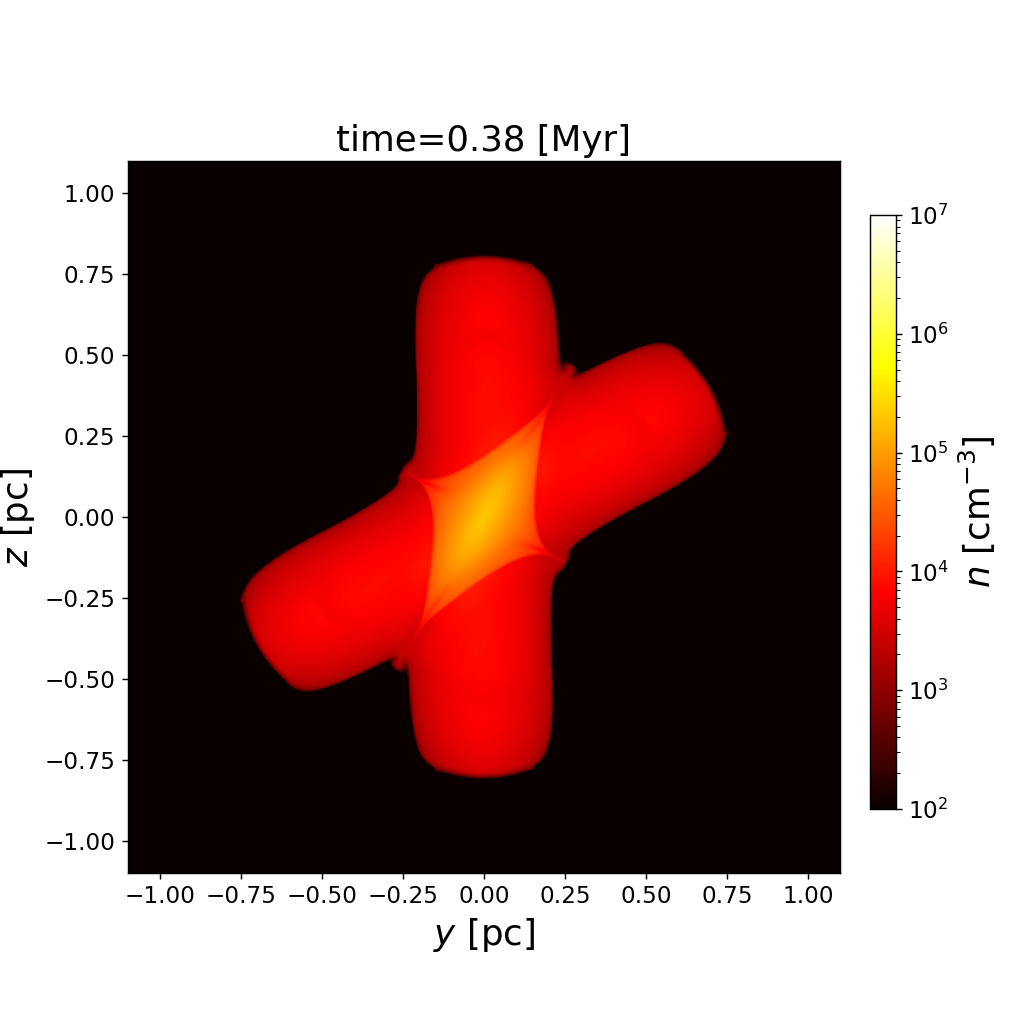}
} &
\subfigure[]{
\includegraphics[width=0.31\textwidth]{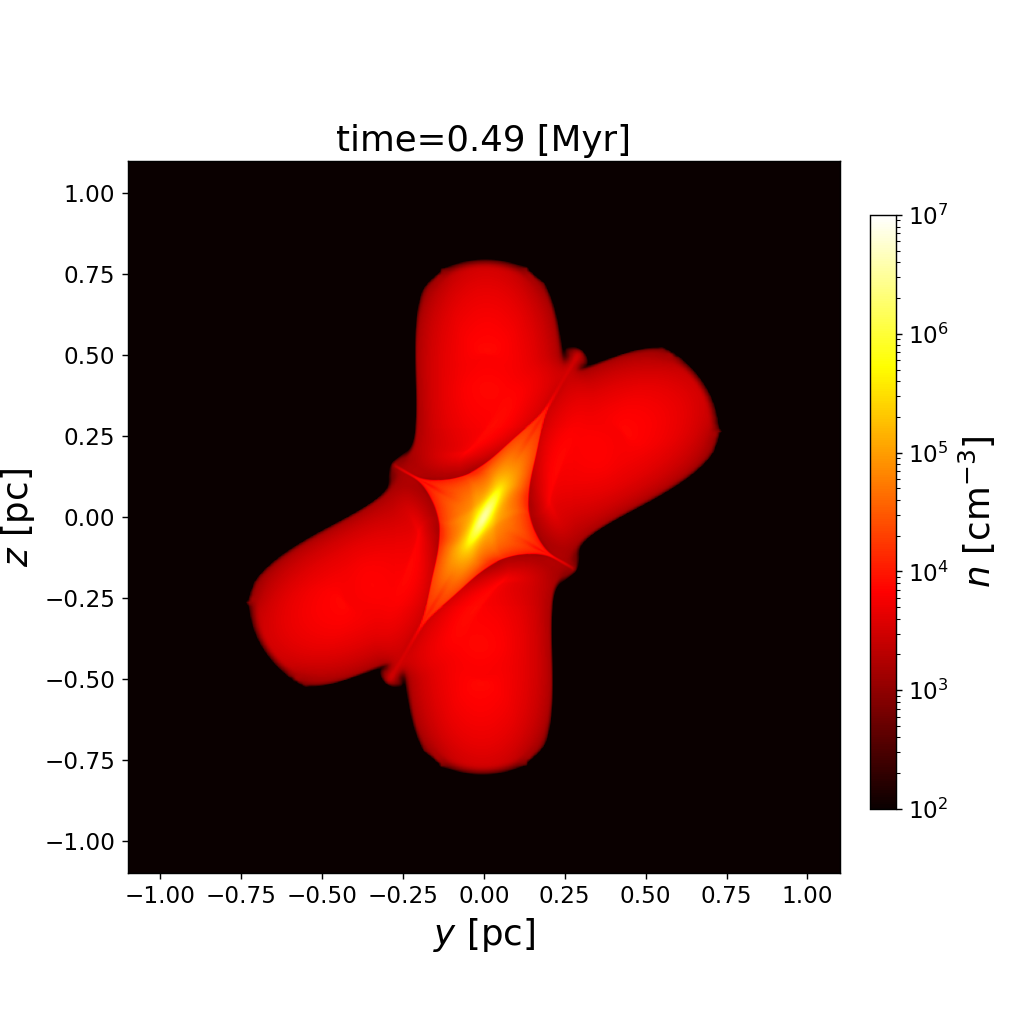}
}
\end{tabular}
\caption{
Evolution of the collapse mode. Two-dimensional slices of the result of model L05V02th60 ($\lambda_0  = 0.5\lambda_{\mathrm{crit,B}}$, $V_\mathrm{int} = c_s$, and $\theta = \pi/3$).
From top to bottom, each row shows slices in the $z = 0$, $y = 0$, and $x = 0$ planes, respectively.
From left to right, the columns correspond to three representative epochs: $t = 0.16\,\mathrm{Myr}$, $0.38\,\mathrm{Myr}$, and $0.49\,\mathrm{Myr}$.
The color scale indicates the density, and the white lines represent the magnetic field lines.
 {Alt text: A set of two-dimensional density slices showing the evolution of the collapse mode for model L05V02th60.}
}
\label{fig:slice_B01L05V02th60}
\end{figure*}

%\begin{comment}
\begin{figure*}[t]
%\centering
\begin{tabular}{ccc}
\subfigure[]{
\includegraphics[width=0.31\textwidth]{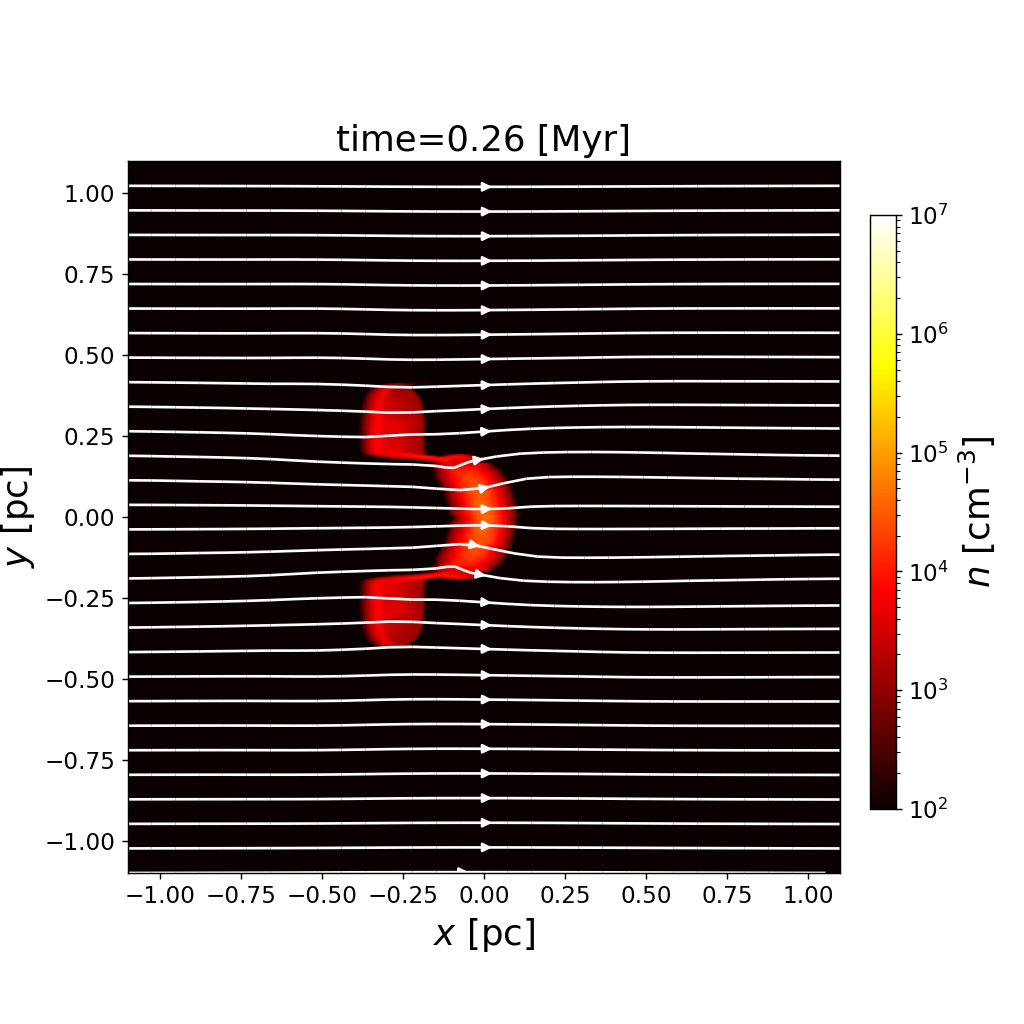}
} &
\subfigure[]{
\includegraphics[width=0.31\textwidth]{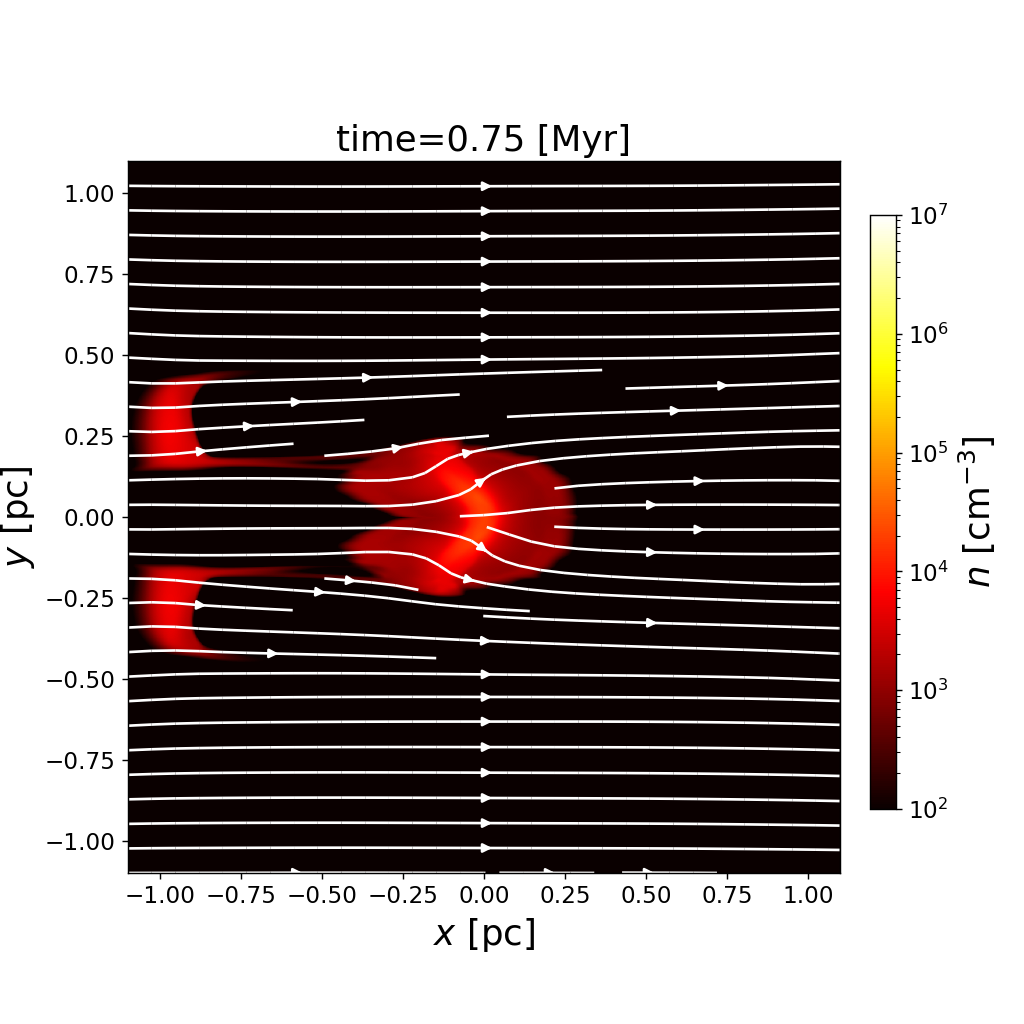}
} &
\subfigure[]{
\includegraphics[width=0.31\textwidth]{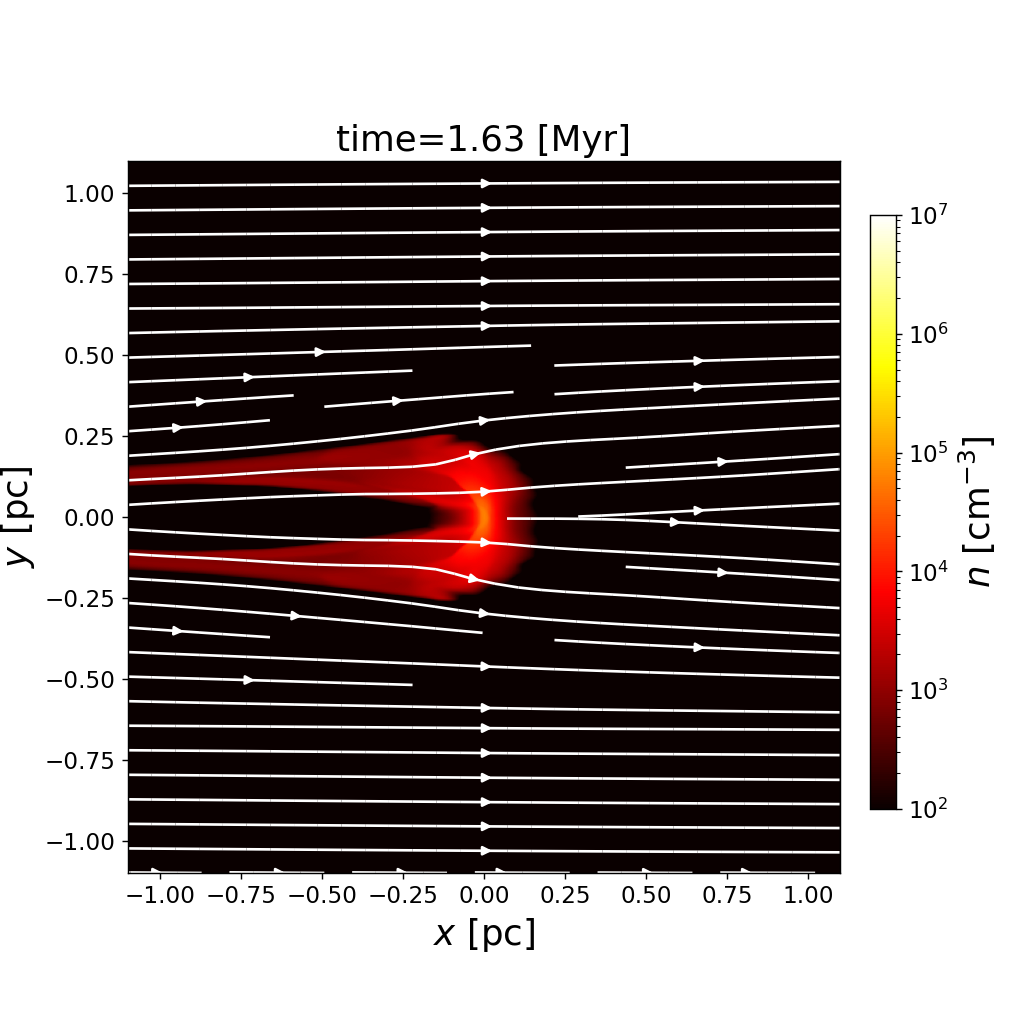}
} \\
\subfigure[]{
\includegraphics[width=0.31\textwidth]{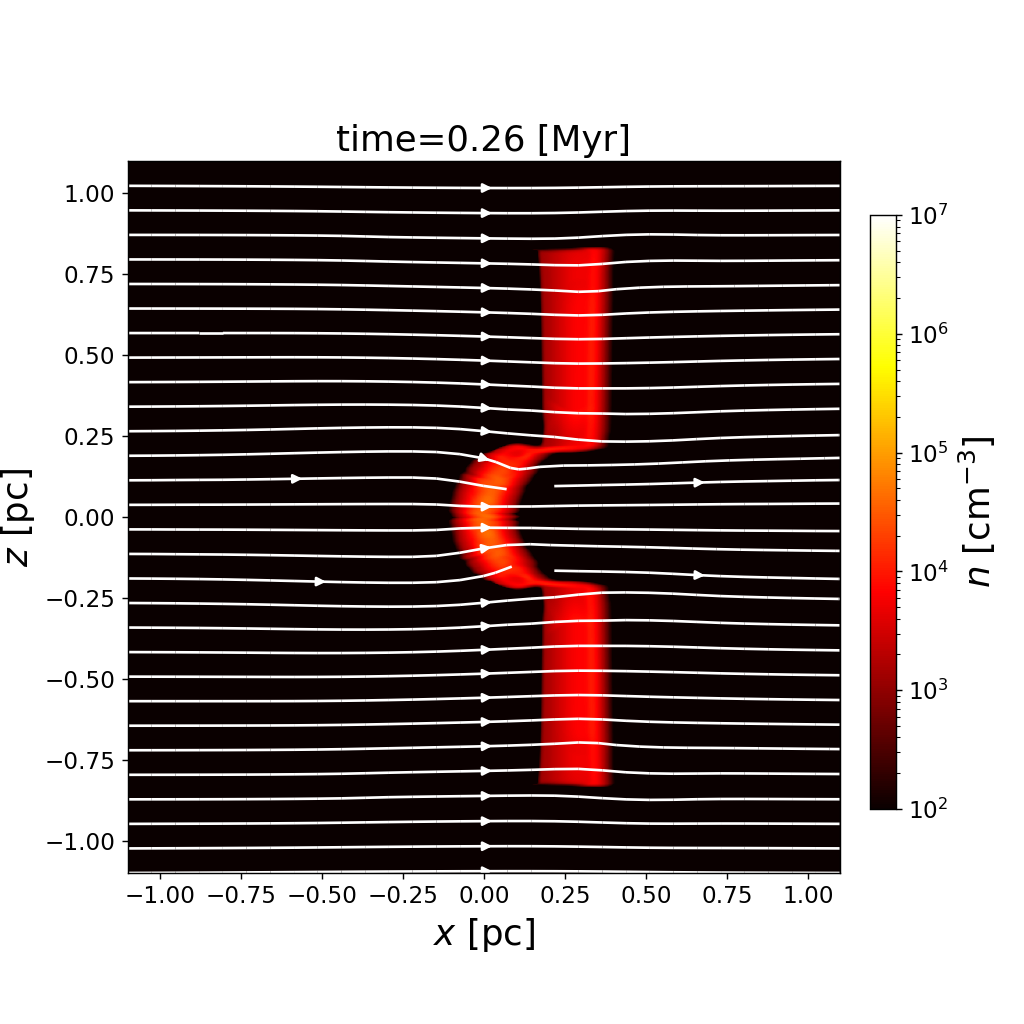}
} &
\subfigure[]{
\includegraphics[width=0.31\textwidth]{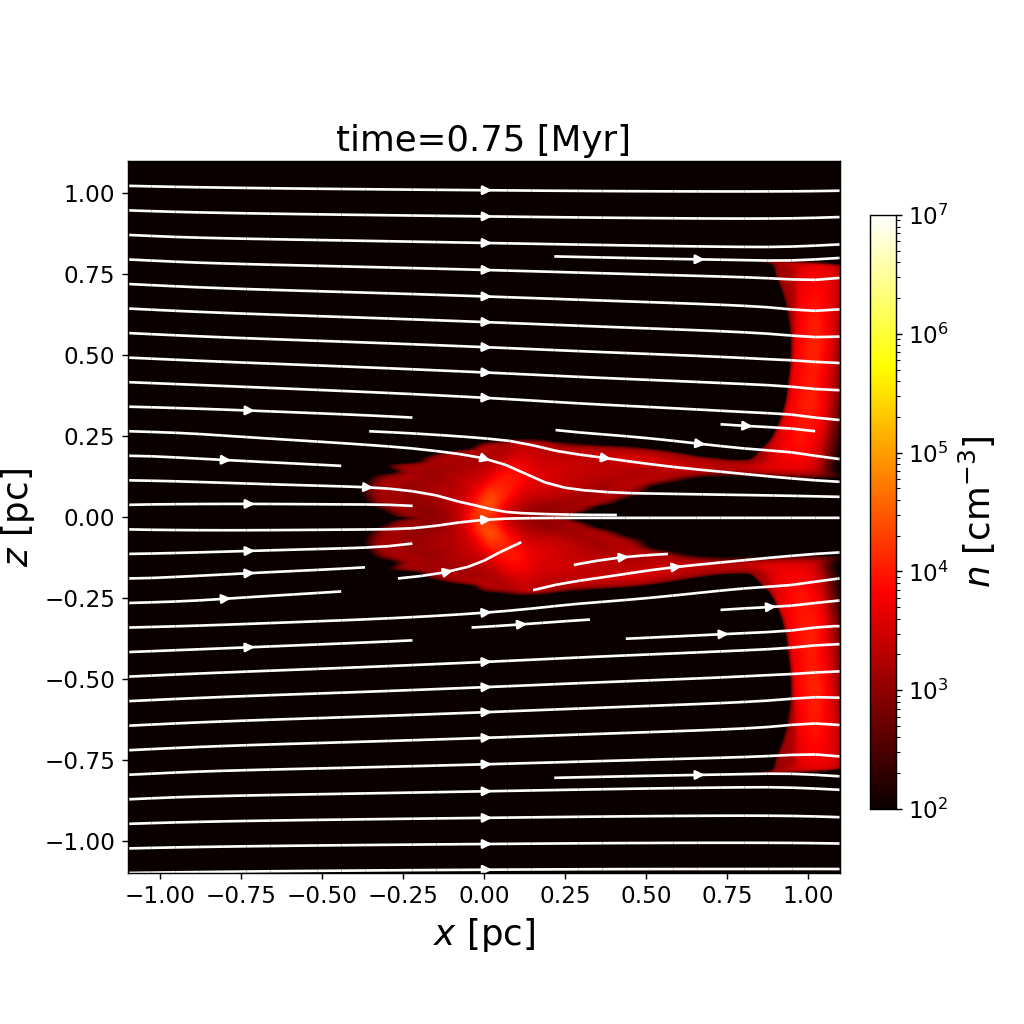}
} &
\subfigure[]{
\includegraphics[width=0.31\textwidth]{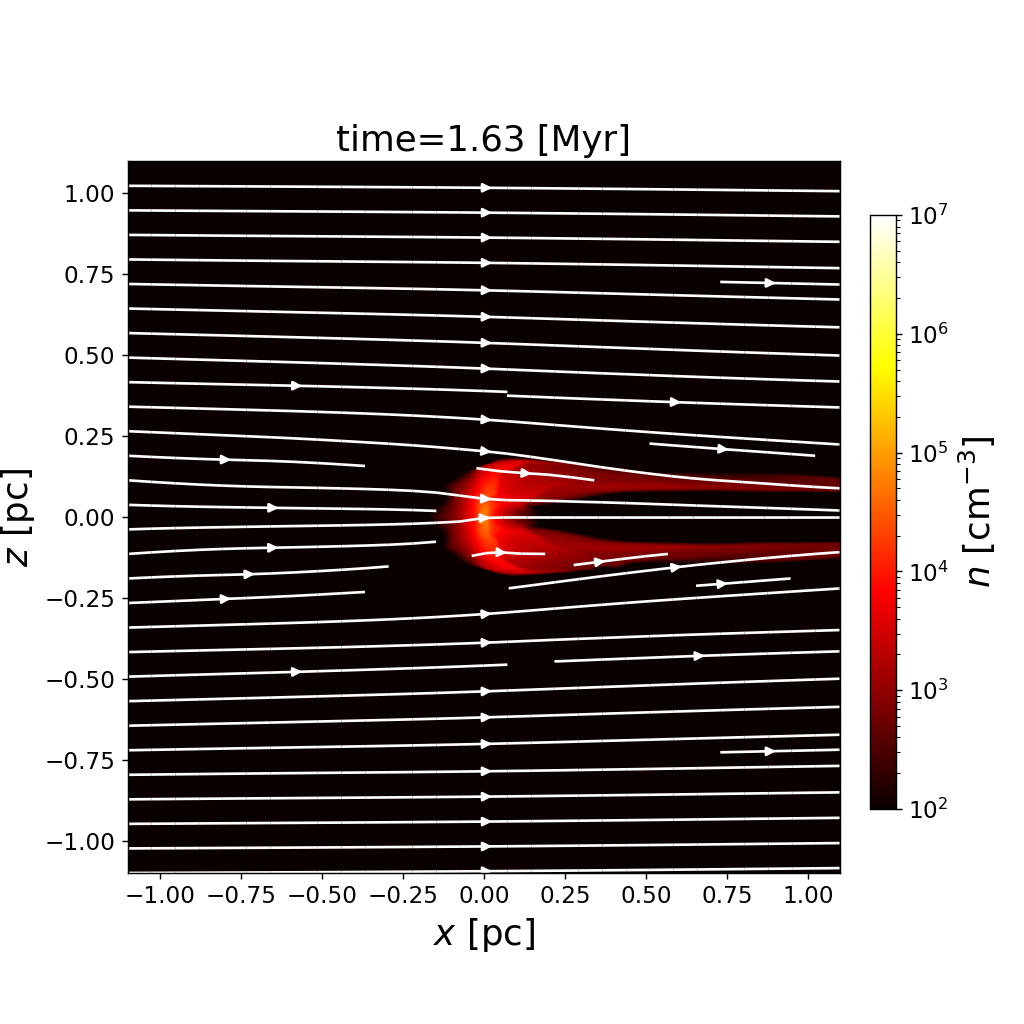}
} \\
\subfigure[]{
\includegraphics[width=0.31\textwidth]{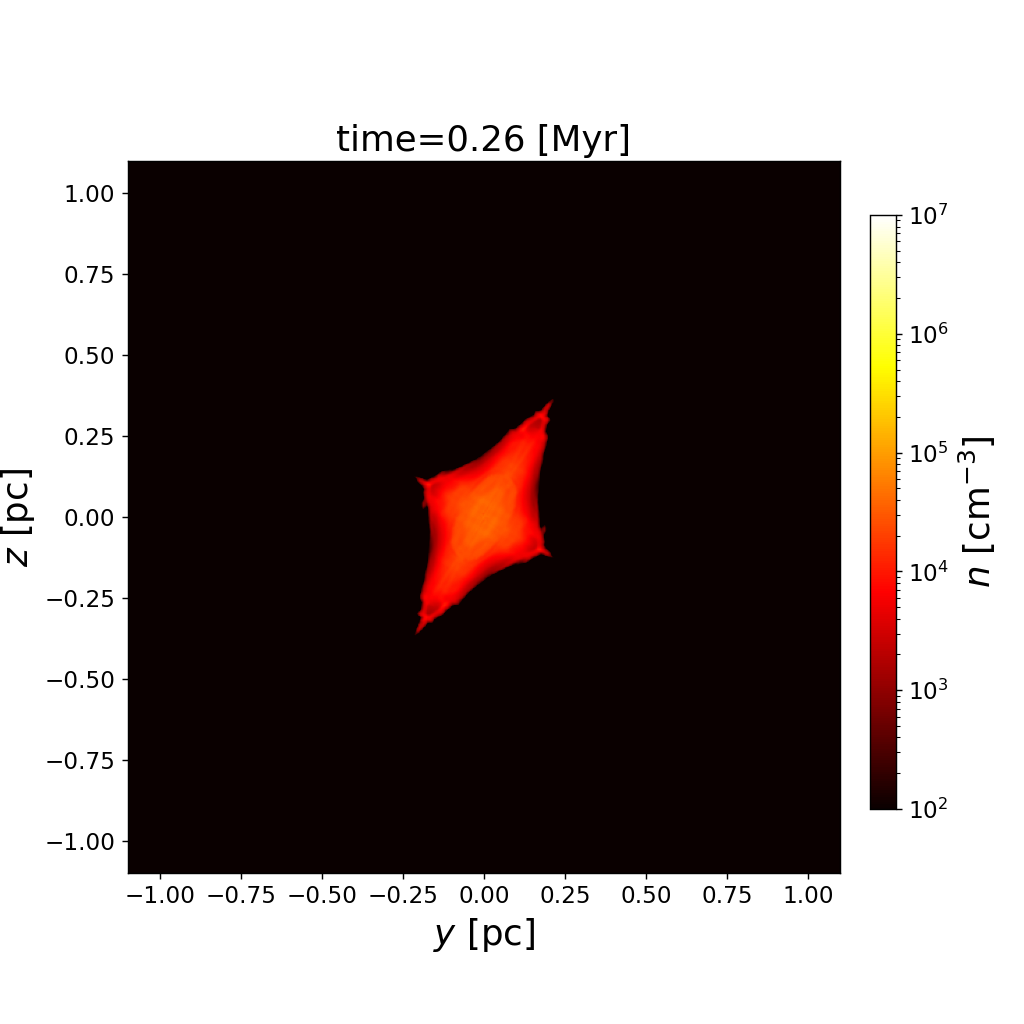}
} &
\subfigure[]{
\includegraphics[width=0.31\textwidth]{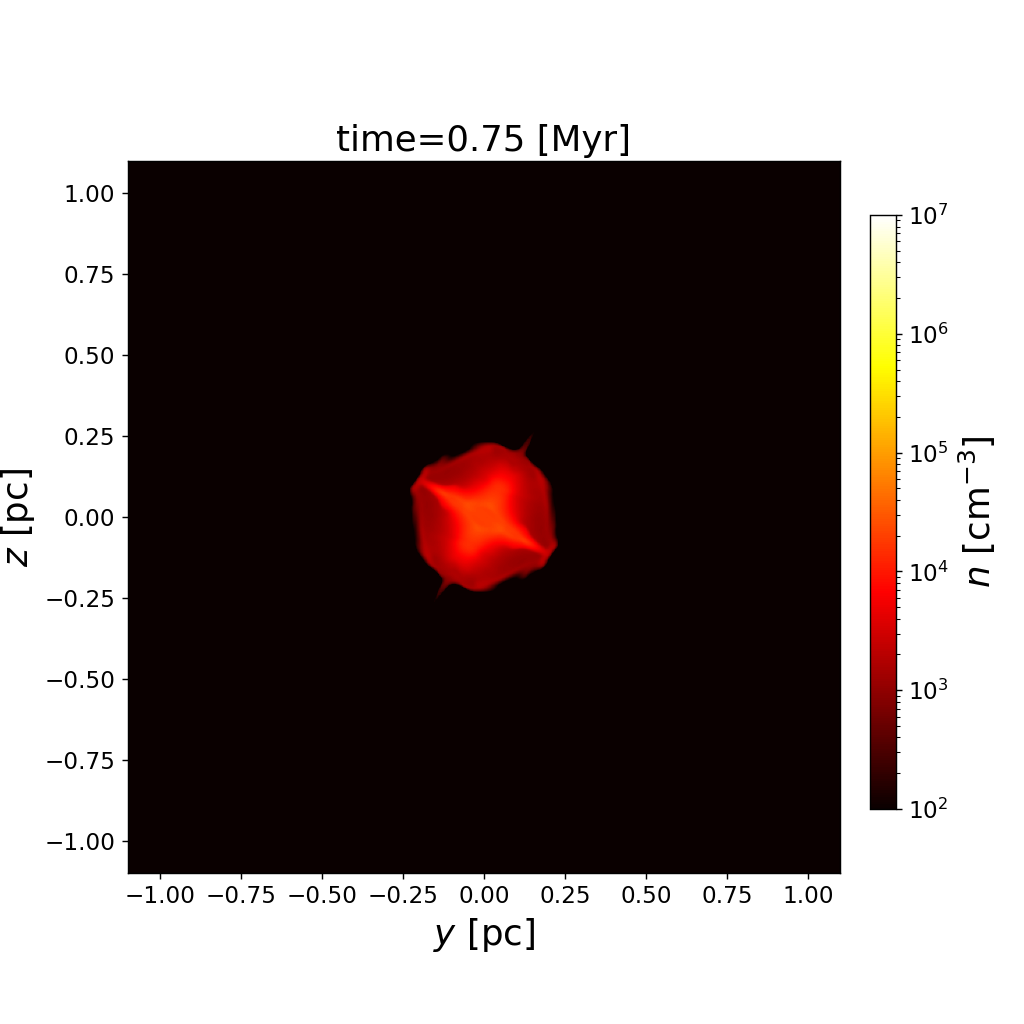}
} &
\subfigure[]{
\includegraphics[width=0.31\textwidth]{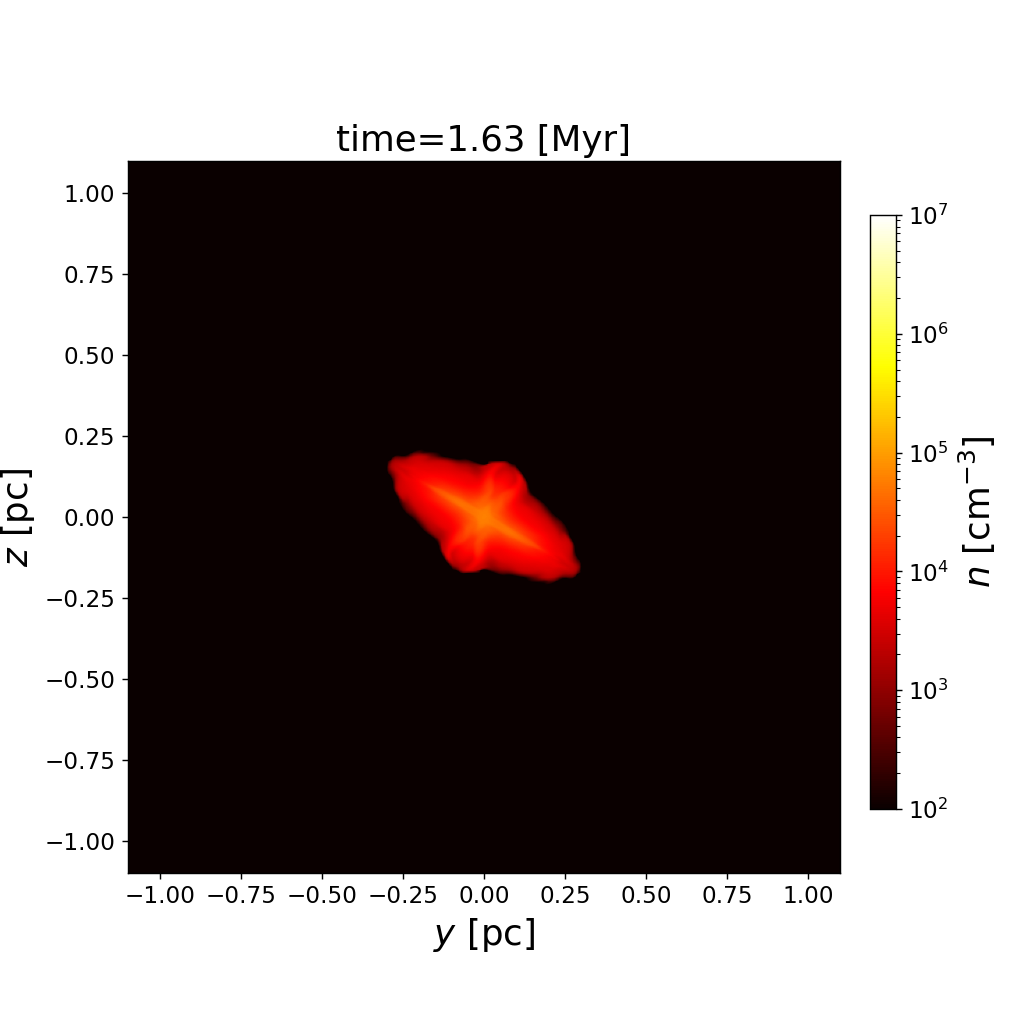}
}
\end{tabular}
\caption{
Same as figure \ref{fig:slice_B01L05V02th60}, but for the expansion mode.
This figure shows results from model L05V16th60
($\lambda_0 = 0.5\,\lambda_{\mathrm{crit,B}}$, $V_\mathrm{int} = 8.0\,c_s$, and $\theta = \pi/3$).
From left to right, columns correspond to three representative epochs:
$t = 0.26\,\mathrm{Myr}$, $0.75\,\mathrm{Myr}$, and $1.63\,\mathrm{Myr}$.
 {Alt text: A set of two-dimensional density slices showing the evolution of the expansion mode for model L05V16th60.}
}
\label{fig:slice_B01L05V16th60}
\end{figure*}

The evolution of oblique collisions can be categorized into two distinct modes: a collapse mode, characterized by a monotonic increase in the density at the intersection, and an expansion mode, in which the intersection expands after the collision. 
In sections \ref{sec:collapse_mode} and \ref{sec:expansion_mode}, we first present the typical evolution of the collapse and expansion modes using fiducial models, respectively.
We then examine the dependence of the resulting evolutionary modes on the model parameters in section \ref{sec:parameter_dep}.

\subsection{Collapse mode}\label{sec:collapse_mode}

In this section, we present the results of the fiducial model for the collapse mode, L05V02th60.
In this paper, the collapse mode is characterized as the case in which the maximum density increases rapidly after the onset of the collision, as shown by the blue solid line in figure \ref{fig:Maxden_vs_time_B01L05th60} for model L05V02th60. 
In this model, the time when the maximum density reaches $n_\mathrm{lim}$ is $t_e = 0.52\,\mathrm{Myr}$ and serves as the fiducial value in determining whether other models belong to the collapse or expansion mode, with all other model parameters being the same (see sections \ref{sec:expansion_mode} and \ref{sec:parameter_dep}).
This definition is based on the fact that the collapse times, measured from the onset of density increase, are nearly identical between the low-velocity collision models ($V_\mathrm{int}=0.5c_s$ and $V_\mathrm{int}=c_s$; see the thick red and blue lines in figure \ref{fig:Maxden_vs_time_B01L05th60}), indicating that $t_e$ in the low-velocity collision model represents the intrinsic collapse timescale independent of the collision velocity. 
Therefore, we adopt $t_e$ of the low-velocity collision model of $V_\mathrm{int}=c_s$ as the fiducial value\footnote{Since $t_e$ in low-velocity collisions varies slightly with the collision angle $\theta$ and the initial line mass $\lambda_0$, we use the value of $t_e$ obtained from the low-velocity collision model corresponding to each $\theta$ and $\lambda_0$ in table \ref{tab:parameters}. }.

Figure \ref{fig:slice_B01L05V02th60} presents the time evolution of two-dimensional slices in the $z=0$ (panels a–c), $y=0$ (panels d–f), and $x=0$ (panels g–i) planes.
At $t = 0.16\,\mathrm{Myr}$, panels (a), (d), and (g) represent the early phase of the collision, during which a shocked region forms at the interface ($x=0$).
In particular, panel (g) reveals that the shocked structure exhibits a rhombus-like shape, with its long axis oriented just between the two filament axes (i.e., $\theta/2$ from the $z$-axis).

At $t = 0.38\,\mathrm{Myr}$, panels (b) and (e) show that the shocked region becomes denser, and the magnetic field lines are dragged toward the center, indicating the onset of global contraction.
Panel (h) represents that the rhombus-like structure becomes more robust, and an elongated dense structure emerges within this region.

At $t = 0.49\,\mathrm{Myr}$, the initial filaments have almost fully collided as shown in panels (c) and (f). 
The central part of the shocked region exhibits further contraction compared to earlier stages, resulting in higher densities and more significant bending of the magnetic field lines.
Panel (i) shows that the elongated dense structure within the rhombus-like shocked region becomes more distinct. 
In oblique collisions, the collapsing dense region tends to form a thin triaxial ellipsoid rather than a disk structure observed in orthogonal collisions (see RK24).

\subsection{Expansion mode}\label{sec:expansion_mode}

In this study, the expansion mode is defined as the case in which either $t_e$, the time required to reach the numerical limit density, or the termination time of the calculation, is longer by a factor of two than $t_e$ of the $V_\mathrm{int} = c_s$ case, with all other model parameters being identical.
This criterion is based on the assumption that, when considering star formation triggered by filament collisions, if the collapse takes too long, it implies that the collision does not significantly contribute to star formation.

In this section, we present the model L05V16th60, adopted as the fiducial case for the expansion mode.
In this model, the shocked region does not undergo runaway collapse even after sufficient time has passed following the collision $t\ge 2t_e(V_\mathrm{int}=c_s)\simeq 1.04\, \mathrm{Myr}$, as shown by the brown line in figure \ref{fig:Maxden_vs_time_B01L05th60}; that is, it remains stable until the termination of the calculation at $t_\mathrm{stop}$.

Figure \ref{fig:slice_B01L05V16th60} shows the time evolution of slices for model L05V16th60. 
At $t = 0.26\,\mathrm{Myr}$, panels (a) and (d) indicate that the collision has already completed, and the non-colliding parts have continued moving past each other along the initial velocity direction.
In addition, panel (g) shows that even in the high-velocity collision case, the shocked region exhibits a rhombus-like shape, similar to that observed in the collapse mode.

At $t = 0.75\,\mathrm{Myr}$, panels (b) and (e) show that the non-colliding parts have nearly reached the numerical boundary, and the shocked region exhibits further expansion compared to the earlier phase.
In panel (h), as the shocked region expands, its shape evolves into a more rounded form.

At $t = 1.63\,\mathrm{Myr}$, panels (c) and (f) show that the non-colliding parts have completely moved beyond the boundary.
Meanwhile, the shocked region has begun to contract, resulting in a slight increase in its density compared to the earlier phase.
In addition, panel (i) shows that the shocked region again takes on a rhombus-like shape, but its major and minor axes are swapped compared to those in panel (g).

\begin{figure*}
    \begin{tabular}{cc}
         \subfigure[]{
         \includegraphics[keepaspectratio,scale=0.4]{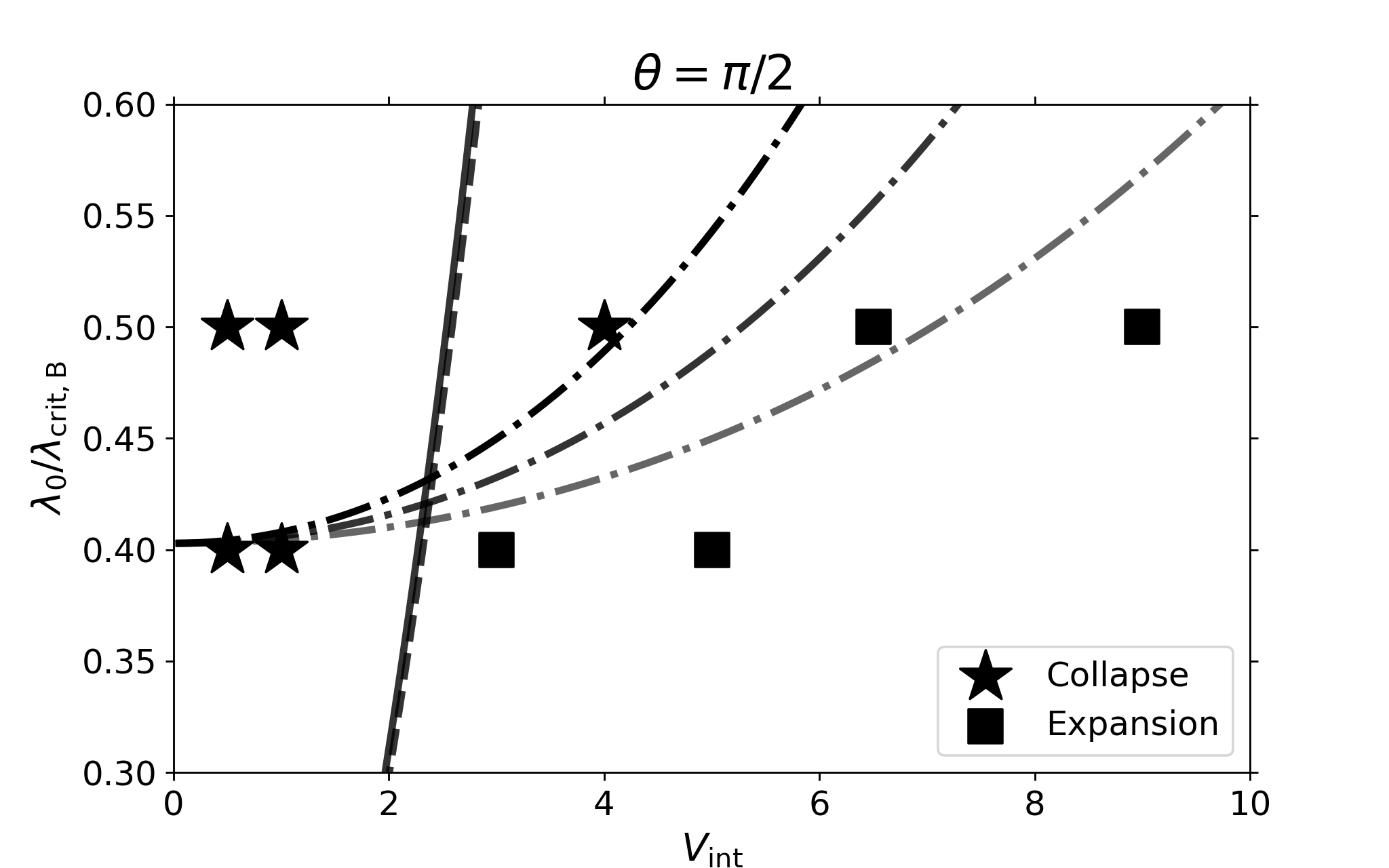}}&
         \subfigure[]{
         \includegraphics[keepaspectratio,scale=0.4]{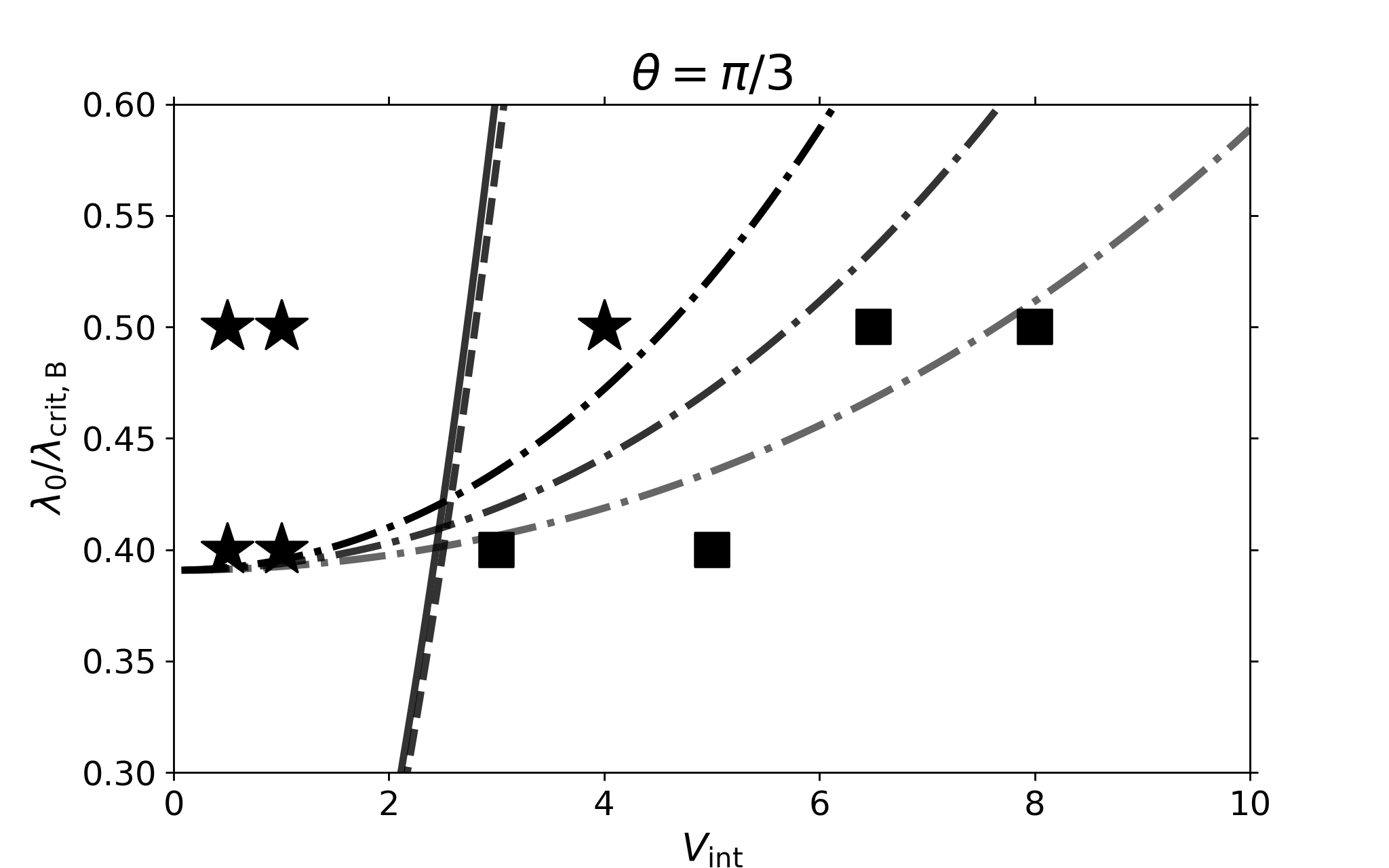}}\\
         &\\
         \multicolumn{2}{c}{\subfigure[]{
         \includegraphics[keepaspectratio,scale=0.4]{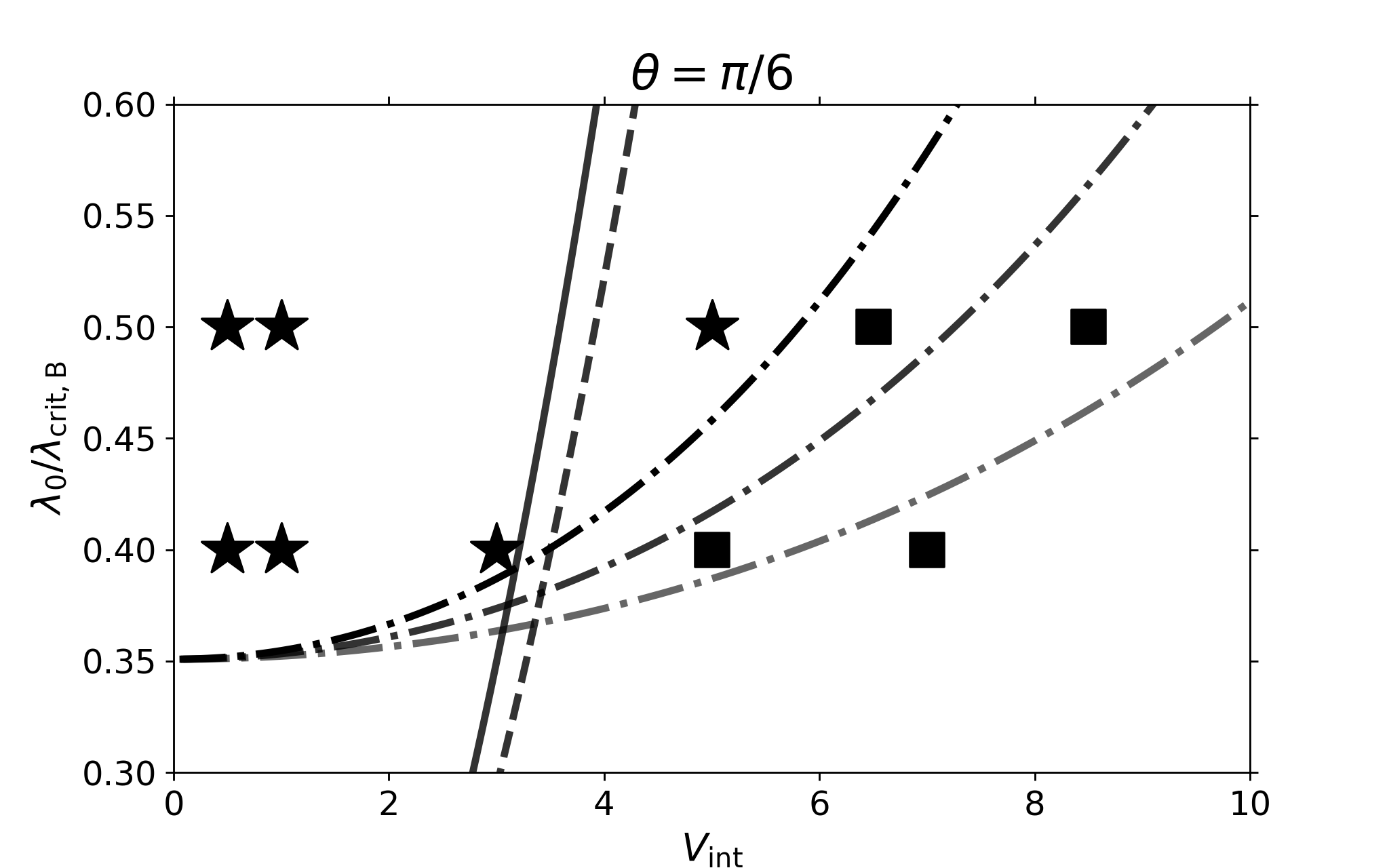}}}
     \end{tabular}
      \caption{
      Final outcomes of oblique collisions, shown on the $V_\mathrm{int}$–$\lambda_0 / \lambda_\mathrm{crit,B}$ plane.
      Panels (a), (b), and (c) correspond to collision angles of $\theta = \pi/2$, $\pi/3$, and $\pi/6$, respectively.
      Star symbols indicate collapse modes, while square symbols represent expansion modes.
      The dash-dotted lines denote the collapse--expansion boundary, representing the threshold line mass (i.e., $\lambda_0 / \lambda_\mathrm{crit,B}$) at which the total energy of the compressed cloud becomes zero immediately after the collision.
      These boundaries are derived from the condition that the total energy equals zero in equation (\ref{eq:KE_after_PE_ME_GE}) in section \ref{sec:condition}, where three models with coefficients of restitution of $\epsilon = 0.5$ (upper), $0.4$ (middle), and $0.3$ (lower) are shown in each panel.
      The solid lines indicate the escape velocity at each line mass for the non-colliding segment of the initial filament from the compressed cloud, as discussed in section \ref{sec:condition_for_escape_velocity}.
      The dashed lines indicate the more accurate escape velocity, as discussed in appendix \ref{sec:Center_of_mass_of_trapezoid}. 
       {Alt text: A set of three panels showing the final outcomes of oblique filament collisions on the initial velocity–line mass plane.}
      }
    \label{fig:stability_for_oblique}
\end{figure*}

\subsection{Dependence of the resulting evolutionary modes on initial parameters}\label{sec:parameter_dep}
In this section, we examine the effect of each parameter on the resulting evolutionary modes. 
Figure \ref{fig:stability_for_oblique} summarizes the evolutionary modes of all models.

Figure \ref{fig:Maxden_vs_time_B01L05th60} clearly shows the effect of the collision velocity; increasing the collision velocity causes the evolutionary mode from collapse to expansion when the other parameters are fixed.
In figure~\ref{fig:Maxden_vs_time_B01L05th60}, the model with $V_\mathrm{int} = 4.0c_s$ (orange line) shows a slight delay but an overall similar $t_e$ to the model with $V_\mathrm{int} = c_s$.
However, the model with $V_\mathrm{int} = 6.5c_s$ (green line) exhibits a noticeably delayed onset of runaway collapse relative to the model with $V_\mathrm{int} = c_s$.
Since the time required to reach the numerical limit density is $t_e = 1.60\,\mathrm{Myr}$, which is more than twice as long as $t_e = 0.52\,\mathrm{Myr}$ for the $V_\mathrm{int} = c_s$ case, we classify the model with $V_\mathrm{int} = 6.5c_s$ (green line) as the expansion mode.
Figure \ref{fig:stability_for_oblique} also shows this transition from collapse to expansion, for collision angles of $\theta = \pi/2$ and $\pi/6$.
In fact, this trend for $\theta = \pi/2$ has already been reported in the case of orthogonal collisions between filaments with infinite length (see RK24).

Figure \ref{fig:stability_for_oblique} indicates that smaller collision angles suppress the emergence of the expansion mode.
Comparing the models with $\lambda_0 = 0.4\lambda_\mathrm{crit,B}$ and $V_\mathrm{int}=3.0 c_s$ for different collision angles $\theta$, we find that the evolutionary mode shifts from expansion to collapse as the collision angle decreases.
This is because a smaller collision angle leads to a more extensive intersection region, which increases the total mass of the shocked region and consequently enhances the gravitational energy (see section \ref{sec:inclination_angle_dependence} for details).

In addition, for a given collision angle, the critical initial velocity at which the evolutionary mode shifts from collapse to expansion increases with line mass.
For example, focusing on the $\theta = \pi/6$ models in figure \ref{fig:stability_for_oblique}, the critical velocity is found to be $V_\mathrm{int} = 3.0$–$5.0c_s$ for $\lambda_0 = 0.4\lambda_\mathrm{crit,B}$, while it increases to $V_\mathrm{int} =$ $5.0$-$6.5c_s$ for $\lambda_0 = 0.5\lambda_\mathrm{crit,B}$.
This indicates that collisions between filaments with higher line mass make the transition to the expansion mode more difficult.
This trend has been reported in the case of orthogonal collisions of infinite filaments (see RK24), and our results confirm that it also holds for oblique collisions.

\begin{figure}
    \includegraphics[keepaspectratio,scale=0.4]{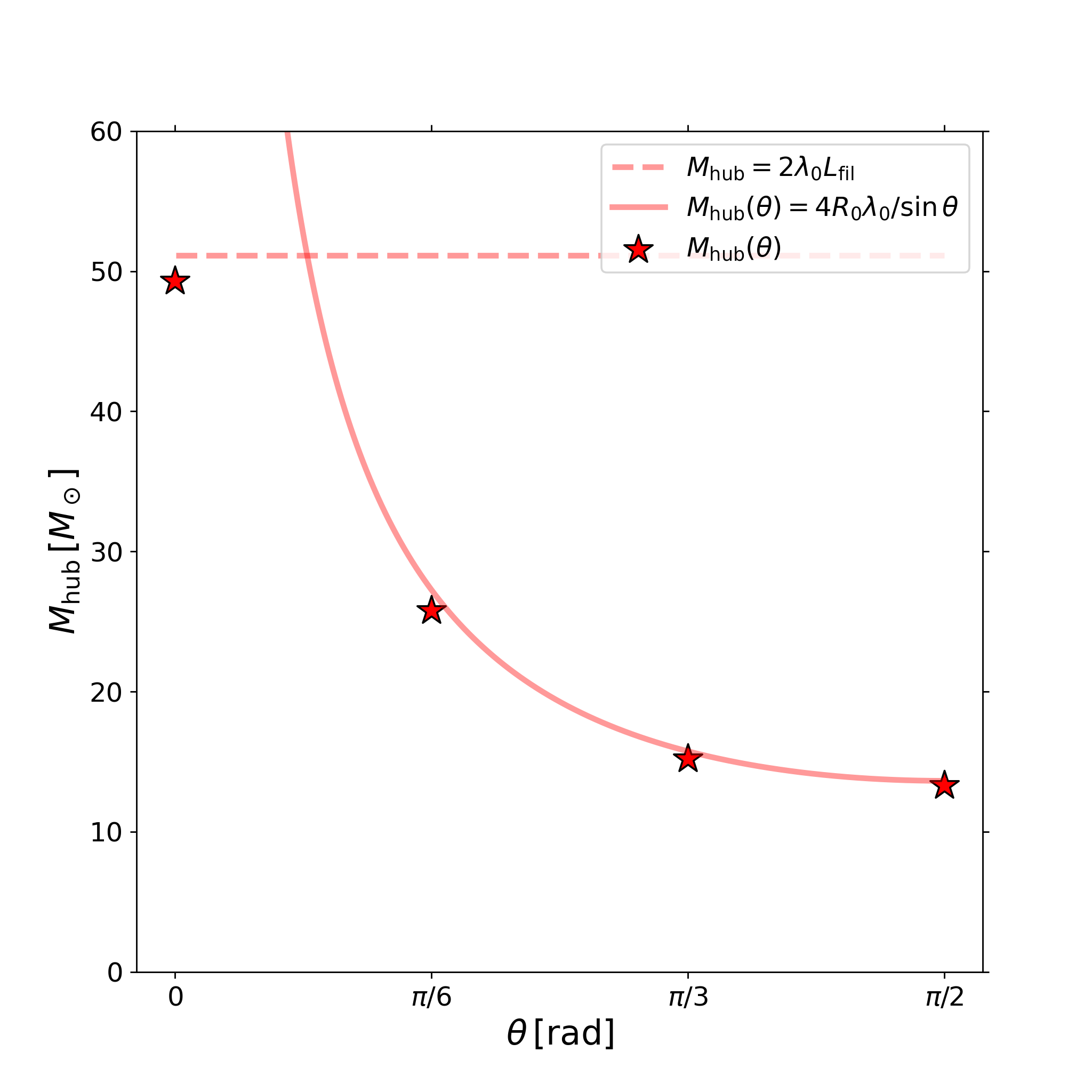}
      \caption{
      Relationship between the collision angle and the mass of the compressed cloud.
      The vertical axis represents the mass of the compressed cloud, and the horizontal axis indicates the collision angle.
      The model parameters are fixed at $\lambda_0  = 0.5\lambda_{\mathrm{crit,B}}$, and $V_\mathrm{int} = 6.5 c_s$, while the collision angle is varied from $\theta=\pi/2$ to $\theta=0$.    
      Star symbols show the mass of gas whose density exceeds the initial central density immediately after the collision.
      The solid curve represents the geometrically estimated compressed mass based on equation (\ref{eq:Delta_M_vs_theta}).
      The dashed line indicates the total mass of two filaments, which gives the upper limit of the compressed gas mass.
      {Alt text: A plot showing the relationship between the collision angle and the mass of the compressed cloud.}
      }
    \label{fig:Delta_M_vs_angle}
\end{figure}

\subsection{Inclination dependence of the mass of the shocked clouds}\label{sec:inclination_angle_dependence}

In this section, we examine the relationship between the collision angle and the mass of the shocked clouds. 
The shocked cloud represents the earliest stage of a hub. 
The amount of gas accumulated at the time of its formation is therefore a key factor 
that determines the subsequent star formation activity within the hub.

In section~\ref{sec:model}, we present a theoretical prediction for the hub mass as a function of the collision angle, showing that a smaller collision angle increases the effective cross-sectional area of the interaction region and thus the mass of the shocked cloud, as given by equation~(\ref{eq:Delta_M_vs_theta}).

To compare with equation (\ref{eq:Delta_M_vs_theta}), we estimate the mass of the shocked cloud in the numerical results as the total mass of gas with densities exceeding the maximum initial density of the filaments, under the assumption that this density enhancement results from compression due to the collision.
The mass is evaluated at the time when the filaments have completely collided \footnote{In this analysis, we also conduct a model with $\theta=0$, but use it only for measuring the mass of the compressed cloud.}.

In figure \ref{fig:Delta_M_vs_angle}, the numerical results show good agreement with the analytic estimate equation (\ref{eq:Delta_M_vs_theta}). 
This agreement indicates that oblique collisions produce more massive shocked clouds compared to orthogonal collisions \footnote{It should be noted that equation (\ref{eq:Delta_M_vs_theta}) is valid for a sufficiently long filament, and $M_\mathrm{hub}$ diverges in the limit as $\theta=0$ as seen in figure \ref{fig:Delta_M_vs_angle}. 
The mass of the shocked cloud formed in collisions of finite-length filaments cannot exceed the total initial filament mass, $2\lambda_0 L_\mathrm{fil}$.}.

\section{Discussion}\label{sec:discussion}

\subsection{Criteria for differentiating collapse and expansion modes}\label{sec:condition}

In the context of induced star formation via filament collisions, stars mainly form from gravitationally collapsing shocked clouds. 
Therefore, determining the conditions under which these clouds undergo collapse is crucial for understanding the initial conditions of star formation.

In this section, we describe the conditions that separate the collapse and expansion modes in the parameter space.
Following RK24, we evaluate the energy balance of the shocked clouds formed immediately after collisions. 
For orthogonal collisions between infinite filaments, RK24 showed that the critical condition for collapse is that the magnitude of the gravitational energy exceeds the sum of the kinetic, thermal, and magnetic energies.
Similarly, in oblique collisions, the subsequent evolution of the compressed cloud can be characterized by using the energy balance of shocked clouds formed immediately after the collisions.

In oblique collisions, the total energy ($E_\mathrm{tot}$) of the shocked cloud immediately after the collision is expressed as %(for clarity, normalized variables are marked with prime)
\begin{equation}\label{eq:KE_after_PE_ME_GE}
\begin{split}
E_\mathrm{tot}&=KE_\mathrm{after}+PE+ME+GE,
%&=\frac{2\lambda'_0 R'_0 \epsilon^2 V'^2_0}{\sin\theta} +\frac{6\lambda'_0 R'_0}{\sin\theta} +\frac{8\eta R'^3_0 B'^2_0}{\sin\theta} \\&   \,\,\,\,\,\,\,\,\,-\frac{12}{5\pi}\frac{R'_0\lambda'^2_0}{\sin^2\theta}\sqrt{2}\sin\left({\theta}/{2}\right)\,F(\alpha).
\end{split}    
\end{equation}
where $KE_\mathrm{after}$, $PE$, $ME$, and $GE$ denote the kinetic, thermal, magnetic, and gravitational energies of the shocked cloud, respectively.

The kinetic energy of the shocked cloud after the collision has fully completed is given by 
\begin{equation}\label{eq:kin_E}
KE_\mathrm{after}= 2\int dV \frac{1}{2}\bar{\rho} V^2_\mathrm{out}\simeq \frac{2\lambda_0 R_0 \epsilon^2 V^2_\mathrm{int}}{\sin\theta},
\end{equation}
where the first factor of 2 comes from the fact that there are two colliding filaments, $\bar{\rho}$ corresponds to the mean density of the initial filament, $\epsilon$ indicates the coefficient of restitution, and $V_\mathrm{int}$ denotes the initial speed of each filament.
Here, we assumed that the expansion velocity after the collision ($V_\mathrm{out}$) is expressed as $V_\mathrm{out}=\epsilon V_\mathrm{int}$.
This $\epsilon$ arises from the fact that a hydrodynamic collision is inelastic, and following RK24, we adopt $\epsilon = 0.4$ as the fiducial value.
Given that the integration range corresponds to the filament volume forming the intersection, $\int dV = \int^{\eta R_0}_{-\eta R_0} \int^{R_0}_{-R_0} \int^{R_0/\sin\theta}_{-R_0/\sin\theta} dx\,dy\,dz$, where $\eta$ is a geometrical factor representing the minor-to-major axis ratio of the initial filament, we fix $\eta = 0.52$, corresponding to the average value at $\beta_0 = 0.1$, and use the relation $\lambda_0 \simeq 4\bar{\rho} \eta R^2_0$.
Similarly, the thermal energy of the shocked cloud is given by
\begin{equation}\label{eq:th_E}
PE = 2\int dV \frac{3}{2}c_s^2\bar{\rho}\simeq\frac{6c^2_s\lambda_0 R_0}{\sin\theta}.
\end{equation}
The magnetic energy of the shocked cloud is given by
\begin{equation}\label{eq:mag_E}
ME = 2\int dV \frac{1}{8\pi}B^2_0\simeq\frac{2\eta R^3_0B^2_0}{\pi\sin\theta},
\end{equation}
where $B_0$ is the uniform magnetic field strength.

The gravitational energy of the shocked cloud is approximated as that of a Jacobi ellipsoid.
This approximation is motivated by the fact that the shocked cloud exhibits a triaxial ellipsoidal shape. 
The Jacobi ellipsoid was originally studied by \citet{1834AnP...109..229J} \citep[see][]{1969efe..book.....C}.
The gravitational energy is given as 
\begin{equation}\label{eq:jacobi}
    GE\simeq - \frac{3}{5}\frac{GM^2}{a_1}\frac{1}{\sin\gamma}\sqrt{2}\int^{\gamma}_{0}\frac{d\phi}{(1-\sin^2{\alpha}\sin^2{\phi})^{1/2}},
\end{equation}
where we use the semi-principle axes, $a_1>a_2>a_3$, of the shocked cloud.
Then, we define $a_2=a_1\cos\beta$, $a_3=a_1\cos\gamma$, therefore $e_{12}=(1-a^2_2/a^2_1)^{1/2}=\sin \beta$, $e_{13}=(1-a^2_3/a^2_1)^{1/2}=\sin \gamma$. 
Using the ratio of these eccentricities, the geometrical parameter is defined as $\alpha \equiv \sin^{-1}(\sin \beta / \sin \gamma)$.
The coefficient of $\sqrt{2}$ arises from the assumption that the compressed cloud is modeled as a triaxial structure inscribed within the collision region.
In this analysis, the relation between the semi-principal axes and the collision angle is given by $a_1 = R_0 / \sin(\theta/2)$ and $a_2 = R_0 / \sin(\pi/2 - \theta/2)$.

Equation (\ref{eq:jacobi}) can be simplified by using the fact that $a_3\ll a_1$, or $\gamma\simeq \pi/2$ as follows:
\begin{equation}\label{eq:GE}
    GE\simeq -\frac{48}{5}\frac{G\lambda^2_0R_0}{\sin^2\theta}\sin\left(\theta/2\right)\sqrt{2}F(\alpha),
\end{equation}
where we use the relation $M\simeq 2\lambda_0\times 2R_0/\sin\theta$, and
\begin{equation}
    F(\alpha)=\int^{\pi/2}_{0}\frac{d\phi}{(1-\sin^2\alpha \sin^2\phi)^{1/2}}.
\end{equation}
Therefore, at $\theta = \pi/2$, equation (\ref{eq:GE}) reduces to the gravitational energy equation derived assuming a Maclaurin spheroid, as in RK24.

Figure \ref{fig:stability_for_oblique} shows that the collapse–expansion boundaries derived from equation (\ref{eq:KE_after_PE_ME_GE}) (i.e., $E_\mathrm{tot}=0$) are in good agreement with the numerical results.
The collapse–expansion boundaries indicate that as the collision angle $\theta$ decreases, the boundaries shift downward overall.
This trend arises because the gravitational energy of the shocked cloud increases as the collision angle $\theta$ decreases [see also equation (\ref{eq:GE})].
For example, when filaments with $\lambda_0 = 0.5\,\lambda_\mathrm{crit,B}$ collide, the critical velocity $V_\mathrm{crit}$ at which the system transitions to the expansion mode is expected from equation (\ref{eq:KE_after_PE_ME_GE}) as $V_\mathrm{crit} = 5.28\,c_s$ for $\theta = \pi/2$, $V_\mathrm{crit} =5.73\,c_s$ for $\theta = \pi/3$, and $V_\mathrm{crit} =7.26\,c_s$ for $\theta = \pi/6$.
This indicates that, as the collision angle decreases from $\theta = \pi/2$ to $\theta = \pi/6$, it becomes more difficult for the compressed cloud to transition to the expansion mode, even at higher collision velocities.
In fact, as shown in figure \ref{fig:stability_for_oblique}, a comparison between the simulation results of models L04V06th90 and L04V06th30 shows that the former exhibits the expansion mode, whereas the latter results in the collapse mode.
However, in panel (c) of figure \ref{fig:stability_for_oblique}, the model with $\theta = \pi/6$, $\lambda_0 = 0.5\lambda_\mathrm{crit,B}$, and $V_\mathrm{int} = 6.5c_s$ exhibits an expansion mode, even though it lies on the collapse side of the collapse--expansion boundary with $\epsilon=0.4$.
This discrepancy arises because $\epsilon$ is fixed at 0.4 in our analysis, whereas in reality, $\epsilon$ tends to approach 0.5 as $\theta$ decreases.
If we instead assume $\epsilon = 0.5$, the theoretical collapse–expansion boundary becomes more consistent with the numerical results.

Although our simulations are performed with $\beta_0 = 0.1$, the collapse–expansion condition for different $\beta_0$ values is derived using equation (\ref{eq:KE_after_PE_ME_GE}). 
Even when the magnetic field strength increases, the trend remains the same, as a smaller collision angle $\theta$ requires a higher collision velocity for the compressed cloud to exhibit the expansion mode.  
However, because the magnetic energy becomes larger, the threshold line mass required for the compressed cloud to undergo gravitational collapse correspondingly increases. 
For completeness, we present this derivation in appendix \ref{sec:property_of_different_beta}.

\subsection{Collision conditions for low-mass hub-filament system formation}\label{sec:condition_for_escape_velocity}

In this section, we propose a collision condition appropriate for the formation of hub–filament systems. 

As shown in figure \ref{fig:slice_B01L05V16th60} (d)–(f), for the model with a high initial velocity, the non-colliding segments have moved out of the computational domain.
This type of evolution is not suitable for hub–filament formation due to the absence of filaments connected to the hub.
Therefore, in order to maintain a hub–filament system, the collision velocity must not exceed a threshold at which the non-colliding segment remains gravitationally bound to the system.
This limit can be reasonably estimated by comparing the initial velocity with the escape velocity from the gravitational potential of the shocked cloud.

The escape velocity $V_\mathrm{esc}$ is defined as 
\begin{equation}\label{eq:escape_vel}
    V_\mathrm{esc}\equiv \sqrt{-2\phi},
\end{equation}
where $\phi$ denotes the gravitational potential of the compressed cloud.
In this study, for simplicity, the compressed cloud is modeled as a sphere. 
In this case, for a point $(x,\,y,\,z)$ located outside the cloud, the escape velocity is expressed as
\begin{equation}\label{eq:escape_vel_hub}
V_\mathrm{esc} = \sqrt{\frac{2GM_\mathrm{hub}}{r}},
\end{equation}
where $r$ denotes the distance between the center of mass of one of the four non-colliding segments and the center of the compressed cloud.
In this analysis, we approximate that the escape velocity evaluated at the center of mass of the non-colliding segment represents the characteristic velocity of the entire segment.

In figure \ref{fig:stability_for_oblique}, the solid lines represent the escape velocities of the non-colliding segments from the compressed cloud, derived for each line mass using equations (\ref{eq:Delta_M_vs_theta}) and (\ref{eq:escape_vel_hub}), where $r$ is fixed at $0.5\mathrm{pc}$.
From panels (a) to (c), the solid lines shift slightly to the right as the collision angle decreases because the increased mass of the shocked cloud deepens the gravitational potential, as seen in section \ref{sec:inclination_angle_dependence}.
Nevertheless, the formation of hub–filament systems becomes unlikely for all collision angles when the initial velocity exceeds approximately two to four times the sound speed within this line mass range.

In reality, the compressed cloud has a triaxial ellipsoidal shape, and the non-colliding segment is trapezoidal. 
A more accurate derivation of the escape velocity is therefore presented in appendix \ref{sec:Center_of_mass_of_trapezoid}.
However, the results obtained using this more accurate treatment (dashed lines in figure \ref{fig:stability_for_oblique}) do not differ significantly from those presented here; therefore, we adopt equation (\ref{eq:escape_vel_hub}) in the analysis.

If the initial velocity exceeds the escape velocity, the formation of a hub–filament system is unlikely, as the non-colliding segments escape from the gravitational potential of the compressed cloud.
Therefore, in figure \ref{fig:stability_for_oblique}, the region above both the dash-dotted and solid lines ideally represents the parameter space favorable for the formation of a hub–filament system resulting from filament collisions. 
An important point is that the solid lines are steeper than the dash-dotted lines, indicating that the escape velocity is insensitive to the initial line mass.
This implies that the escape velocity is smaller than the critical velocity given by the collapse--expansion condition when $\lambda_0\gtrsim0.4\lambda_\mathrm{crit,B}$.
Therefore, in the context of hub-filament formation, we expect that the escape velocity is more important than the critical velocity.

For completeness, the escape velocities for different values of $\beta_0$ are also derived in appendix \ref{sec:property_of_different_beta}.

\subsection{Implications for the formation of massive-star-forming hub-filament systems}\label{sec:condition_for_escape_velocity_for_massive}

We next extend our discussion to more massive hub formation, with total masses in the range of $10^2$–$10^3\,M_\odot$, comparable to those observed in candidate regions of massive star formation \citep[e.g.,][]{2013ApJ...766..115K,2014A&A...561A..83P,2017A&A...606A.123H,2023ApJ...953...40Y,2023ApJ...950..148M}.

We consider the formation of a massive hub at an early evolutionary stage.
However, as shown in table~\ref{tab:parameters}, the maximum hub mass formed in the collision of low-mass filaments in our simulations is $M_\mathrm{hub}=27.3\,M_\odot$, which is insufficient for the formation of a massive hub.
Although the hub mass increases with decreasing collision angle (section~\ref{sec:inclination_angle_dependence}), the total mass of the two low-mass filaments remains below $100\,M_\odot$.
Therefore, collisions between massive filaments ($\gtrsim 100\,M_\odot\,\mathrm{pc^{-1}}$) are more favorable for forming a massive hub at an early evolutionary stage, as they naturally yield a large hub mass from the outset.

One possible formation mechanism for such massive filaments is cloud-cloud collisions \citep{2013ApJ...774L..31I,2021PASJ...73S...1F}, in which shock compression forms a dense sheet, and subsequent converging flows along the sheet assemble massive filaments; see \citet{2018PASJ...70S..53I,2025ApJ...987..154A} for details.
Observationally, massive filaments are known to exhibit turbulent motions, with the total velocity dispersion given by 
$\sigma_\mathrm{tot}^2 = c_s^2 + \sigma_\mathrm{nt}^2$, where $c_s$ and $\sigma_\mathrm{nt}$ correspond to the thermal and non-thermal components, respectively, and typically $\sigma_\mathrm{tot}/c_s \gtrsim 2$ \citep[e.g., see figure 6 of][]{2023ASPC..534..153H}. 
In addition, massive filaments are also threaded by strong magnetic fields, $B_0 \sim 200\,\mu\mathrm{G}$ \citep{2021A&A...647A..78A}.
Including turbulent support, the magnetic critical line mass \citep{2014ApJ...785...24T} becomes
\begin{equation}\label{eq:magnetized_linemass_with_turb}
\begin{split}
    \lambda_\mathrm{crit,B}\simeq &177.9\left(\frac{R_0}{0.2\,\mathrm{pc}}\right)\left(\frac{B_0}{200\,\mu\mathrm{G}}\right)
    \\&+69.7\left(\frac{c_s}{190\,\mathrm{m\,s^{-1}}}\right)^{2}\left(\frac{\sigma_\mathrm{tot}/c_s}{\sqrt{5}}\right)^2M_\odot\,\mathrm{pc^{-1}}.
\end{split}
\end{equation}
For example, adopting $R_0=0.2\, \mathrm{pc}$, $B_0=200\,\mu \mathrm{G}$, $c_s=190\,\mathrm{m\,s^{-1}}$, and $\sigma_\mathrm{tot}/c_s=\sqrt{5}$, we obtain $ \lambda_\mathrm{crit,B}\simeq 250\,M_\odot \,\mathrm{pc^{-1}}$.
Thus, when supersonic turbulent support and strong magnetic fields are included, such filaments can sustain line masses of $\sim 10^2\,M_\odot\,\mathrm{pc^{-1}}$ and can therefore serve as progenitors of massive hubs.

We note that the treatment of turbulence through an effective sound speed in equation (\ref{eq:magnetized_linemass_with_turb}) is intended to approximately describe situations in which additional turbulent support increases the critical line mass and allows even massive filaments to avoid global gravitational collapse.
In reality, supersonic turbulence is expected to generate anisotropic density and velocity structures while also distorting the magnetic field geometry before the collision, potentially altering the collision dynamics and the resulting hub structure.
Nevertheless, if the filamentary structure is approximately maintained and the collision is driven by coherent large-scale motions, we expect that the underlying energy-balance interpretation for the collapse-expansion condition remains qualitatively valid.
Similarly, regarding the escape velocity, even in such environments, there should still exist conditions under which non-colliding segments may remain gravitationally connected to the shocked region. 
Thus, these conditions still provide useful guidelines for understanding massive hub formation.

If massive filaments with these properties collide, the resulting hubs are expected to reach masses of $10^2$–$10^3\,M_\odot$. 
For example, by substituting a line mass of $\lambda_0=250\,M_\odot\,\mathrm{pc^{-1}}$ and $R_0 = 0.2\,\mathrm{pc}$ into equation (\ref{eq:Delta_M_vs_theta}), we obtain a hub mass of $\sim 200\,M_\odot$ even for the orthogonal collision case corresponding to the smallest collision area.
In an oblique collision with $\theta=\pi/6$, the hub mass increases to $\sim 400\,M_\odot$ for the same $R_0$.
Thus, collisions between massive filaments with small inclination angles can form more massive hubs.

For such massive hubs, the relation between the escape velocity and the hub mass is derived from equation (\ref{eq:escape_vel_hub}) as
\begin{equation}\label{eq:escape_ex}
V_\mathrm{esc} \simeq 1.3 \sqrt{\left(\frac{M_\mathrm{hub}}{100\,M_\odot}\right) \left(\frac{r}{0.5\,\mathrm{pc}}\right)^{-1}} \, \mathrm{km\,s^{-1}}. 
\end{equation} 
Since the escape velocity depends not only on $M_\mathrm{hub}$ but also on the distance between the center of mass of the non-colliding segment and the center of the compressed cloud $r$, $V_{\mathrm{esc}}$ becomes smaller as $r$ increases.
Assuming that the filament length is a few pc, we vary the distance $r$ between $0.25$ and $1.0$~pc for reference.
In that case, the collision velocity should be limited to about $1$–$2~\mathrm{km\,s^{-1}}$ for $M_\mathrm{hub}\sim100~M_\odot$, and $3$–$6~\mathrm{km\,s^{-1}}$ for $M_\mathrm{hub}\sim1000~M_\odot$.
If the hub mass, the length of the non-colliding filament segment, and its velocity are known, one can roughly determine whether a hub–filament system can form.
Therefore, equation (\ref{eq:escape_ex}) may serve as useful diagnostics for identifying objects in which hubs are formed via filament merging.

\section{Summary and future prospects}\label{sec:summary}

We conduct three-dimensional ideal MHD simulations to investigate oblique collisions between two identical, finite-length filaments embedded in a lateral magnetic field. 
The filaments share a common magnetic flux tube, and the collision occurs along the direction parallel to the global magnetic field. 
By varying the angle between the filament axes, the initial velocity, and its line mass, we explore a total of 28 models, as summarized in table \ref{tab:parameters}.
Our findings are summarized as follows.

\begin{enumerate}
\item In oblique collisions, unlike orthogonal ones, the compressed cloud takes on a rhombus-like shape. 
Subsequently, the cloud exhibits either collapse or expansion modes. 
As the collision angle approaches zero, the mass contained in the compressed cloud increases, making the cloud easier to collapse even under identical conditions except for the collision angle.
In other words, the critical velocity at which the evolutionary mode transitions from collapse to expansion modes increases as the collision angle decreases.

\item The criterion for distinguishing between collapse and expansion modes can be explained by evaluating the energy balance within the compressed cloud immediately after the collision is complete. 
We derived a condition that the cloud undergoes gravitational collapse if its gravitational energy exceeds the sum of the other energy components; otherwise, it exhibits expansion. We also confirmed that this criterion is consistent with the results of our numerical simulations.

\item We proposed an upper limit on the collision velocity as a necessary condition for the formation of a hub–filament system through filament collisions. 
If the initial velocity is excessively high, the non-colliding segments retain their initial momentum and pass through the compressed region without being gravitationally bound. 
That is, hub–filament systems are unlikely to form when the initial collision velocity exceeds the escape velocity derived from equation (\ref{eq:escape_ex}).
This equation is considered a useful diagnostic for identifying whether observed hubs have formed through filament collisions.

\end{enumerate}

By combining the present study with our previous work \citep{2023ApJ...954..129K,2024ApJ...974..265K}, we have obtained a more comprehensive understanding of the evolution and instability of compressed clouds formed through filament collisions.
Based on our current results, we suggest that collisions between massive filaments, occurring at small collision angles and with velocities below the escape speed, are favorable for forming hub–filament systems associated with massive star formation. 
In addition, according to theoretical studies of filament formation in shock-compressed layers \citep[e.g.,][]{2013ApJ...774L..31I,2021ApJ...916...83A}, filaments are created within shocked sheets, suggesting that small-angle collisions occur much more naturally than randomly distributed ones.
Therefore, further investigations focusing on nearly parallel collisions will be essential in future studies to examine whether such collisions can reproduce massive hubs.

Our simulations follow the evolution up to the onset of gravitational collapse, and no gas accretion begins along the filaments at this stage.
This accretion along the filaments is thought to play an important role in the mass accumulation onto the hub, and observations have suggested gas accretion or bulk flows toward hub regions \citep[e.g.,][]{2013A&A...555A.112P,2014A&A...561A..83P,2013ApJ...766..115K,2017A&A...602L...2H,2018ApJ...852...12Y,2020ApJ...903...13D,2023ApJ...958...17P,2024MNRAS.528.2199R}.
We considered a model in which these accretion flows are driven by the gravity of the hub itself, as discussed in section \ref{sec:condition_for_escape_velocity}.
In such cases, we expect that a more massive hub would promote stronger accretion owing to its deeper gravitational potential.
To further advance our understanding of star formation triggered by filament collisions, it is crucial to perform longer-term simulations.
Such simulations will allow us to verify whether filament collisions can indeed lead to the formation of massive stars and to investigate the subsequent evolution of hub–filament systems.
In addition, long-term simulations enable us to test the validity of the escape velocity proposed in this study.
While this work provides a theoretical prediction of a collision condition for this scenario, a more detailed investigation will be presented in a forthcoming study.

%%%%%%%%%%%%%%%%%%%%%%%%%%%%%%%%%%%%%%% 

% See the instraction below for "Alt text"
% https://academic.oup.com/pasj/pages/General_Instructions#Figures%20and%20Illustrations

%%%%%%%%%%%%%%%%%%%%%%%%%%%%%%%%%%%%%%%

%\section*{Supplementary data} 

%The following supplementary data is available at PASJ online.

%E-table 1  

\begin{ack}
We are grateful to the anonymous referee for careful reading of the manuscript and for insightful comments and suggestions that significantly improved this work.
We thank T. Hoang, R.S. Furuya, M.N. Machida, K. Tokuda, D. Abe, R. Maeda, and S. Higashi for valuable discussions.
Numerical calculations were performed in Yukawa-21 at the Yukawa Institute Computer Facility and Cray XD 2000 at the Center for Computational Astrophysics, National Astronomical Observatory of Japan.
\end{ack}

\section*{Funding}
This work was supported by JSPS KAKENHI Grant Numbers JP22J11106, JP25K17443 (RK), JP21H04487, JP22KK0043 (KI), JP19K03919 (KT), and JP23H00129 (TI).

%\section*{Data availability} 
% The data underlying this article are available ...  
% Sample Data Availability Statements 
% https://academic.oup.com/pages/open-research/research-data#Data%20Availability%20Statements
\bibliographystyle{aasjournal}
\bibliography{accretion_ffc}{}

\appendix %%%%%%%%%%%%%%%%%%%%%%%%%%%%%%%%%%%%%%%%%%%%%%%%%%%%%%%%
%\section*{Case of single paragraph}
% No section number is necessary. Add ``*'' after \verb/\section/.

%%%% 

\section{Dependency of the collapse–expansion boundary and the escape velocity on the plasma beta}\label{sec:property_of_different_beta} 

\begin{figure*}[t]
%\centering
\begin{tabular}{ccc}
\subfigure[]{
\includegraphics[width=0.31\textwidth]{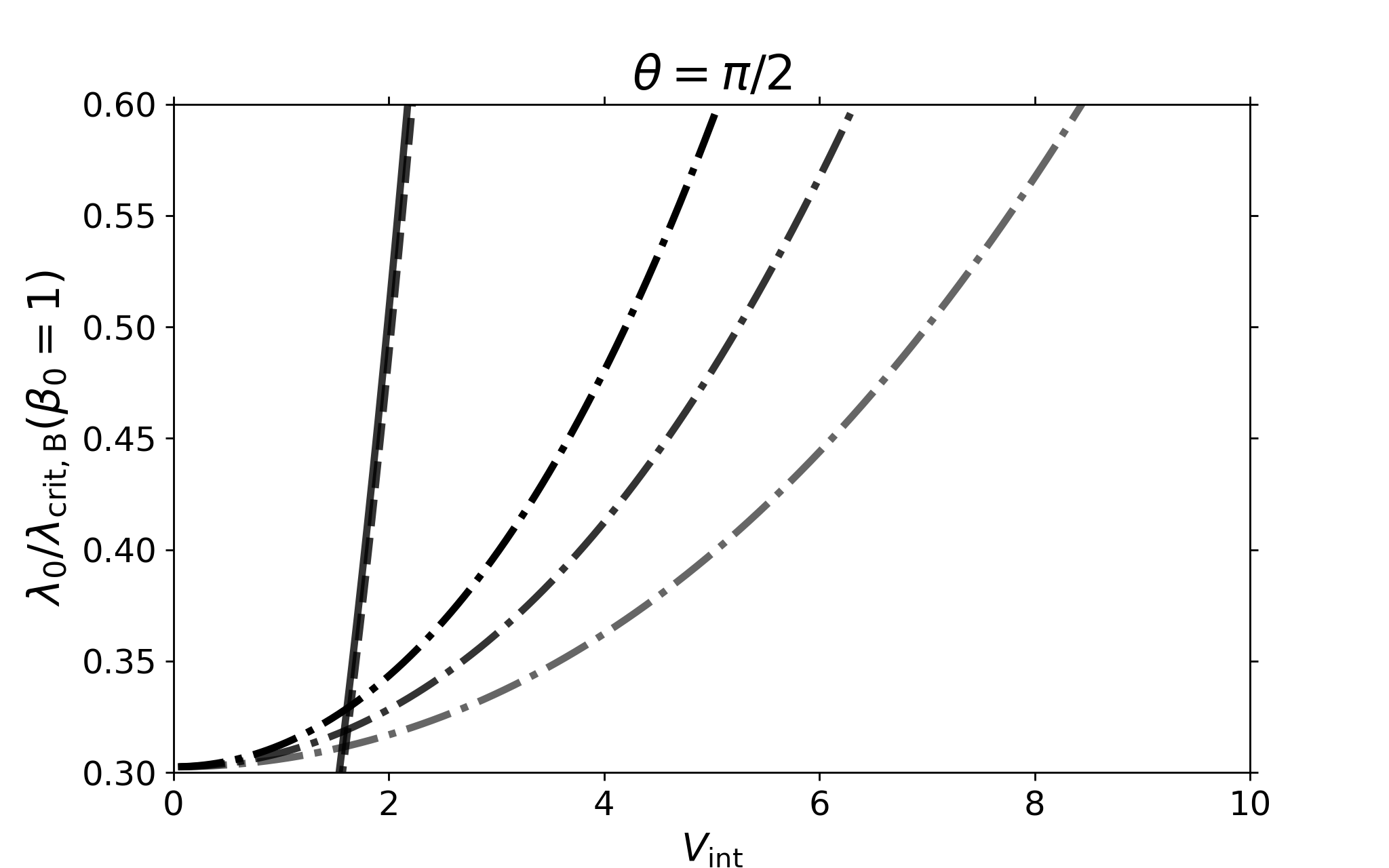}
} &
\subfigure[]{
\includegraphics[width=0.31\textwidth]{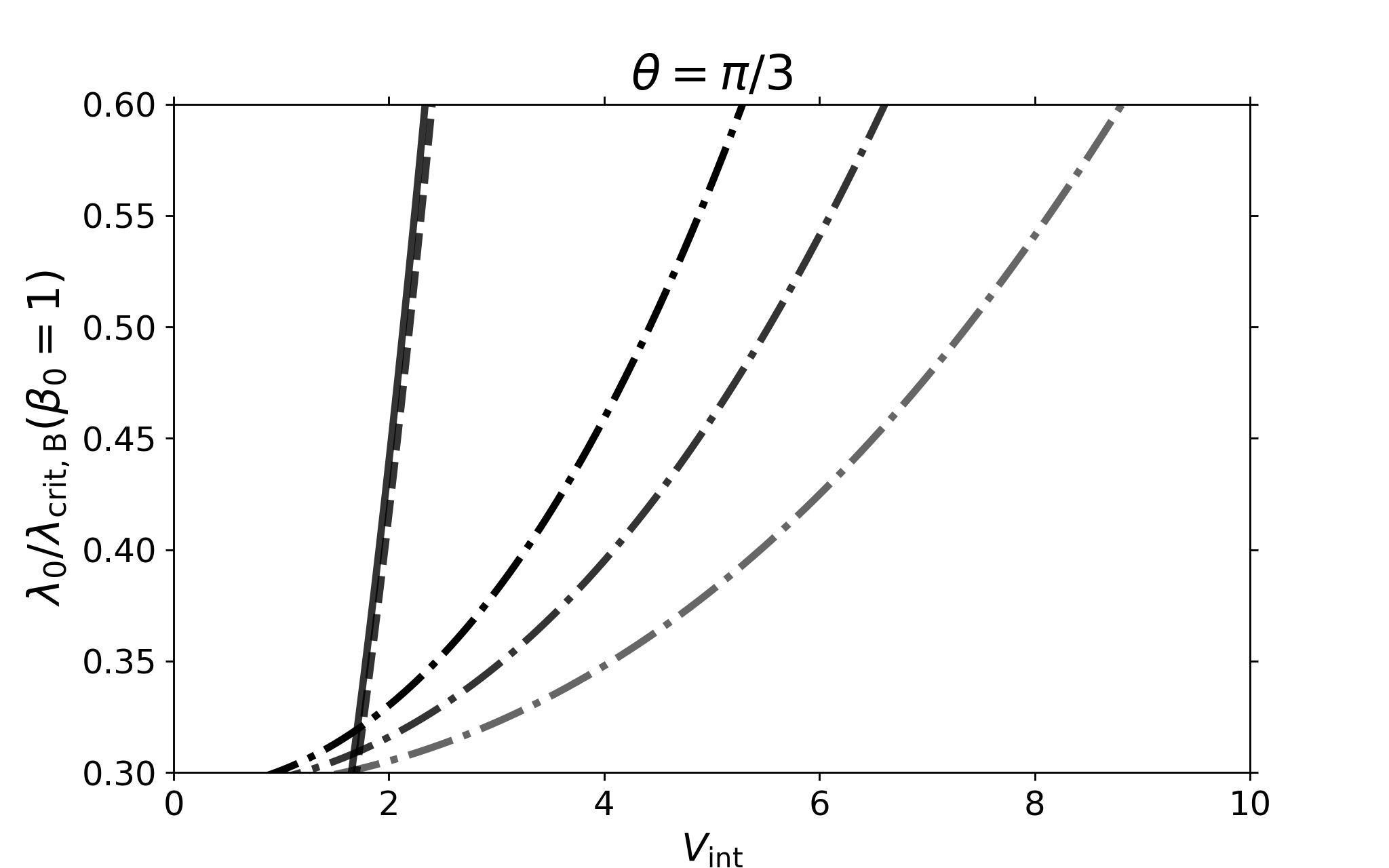}
} &
\subfigure[]{
\includegraphics[width=0.31\textwidth]{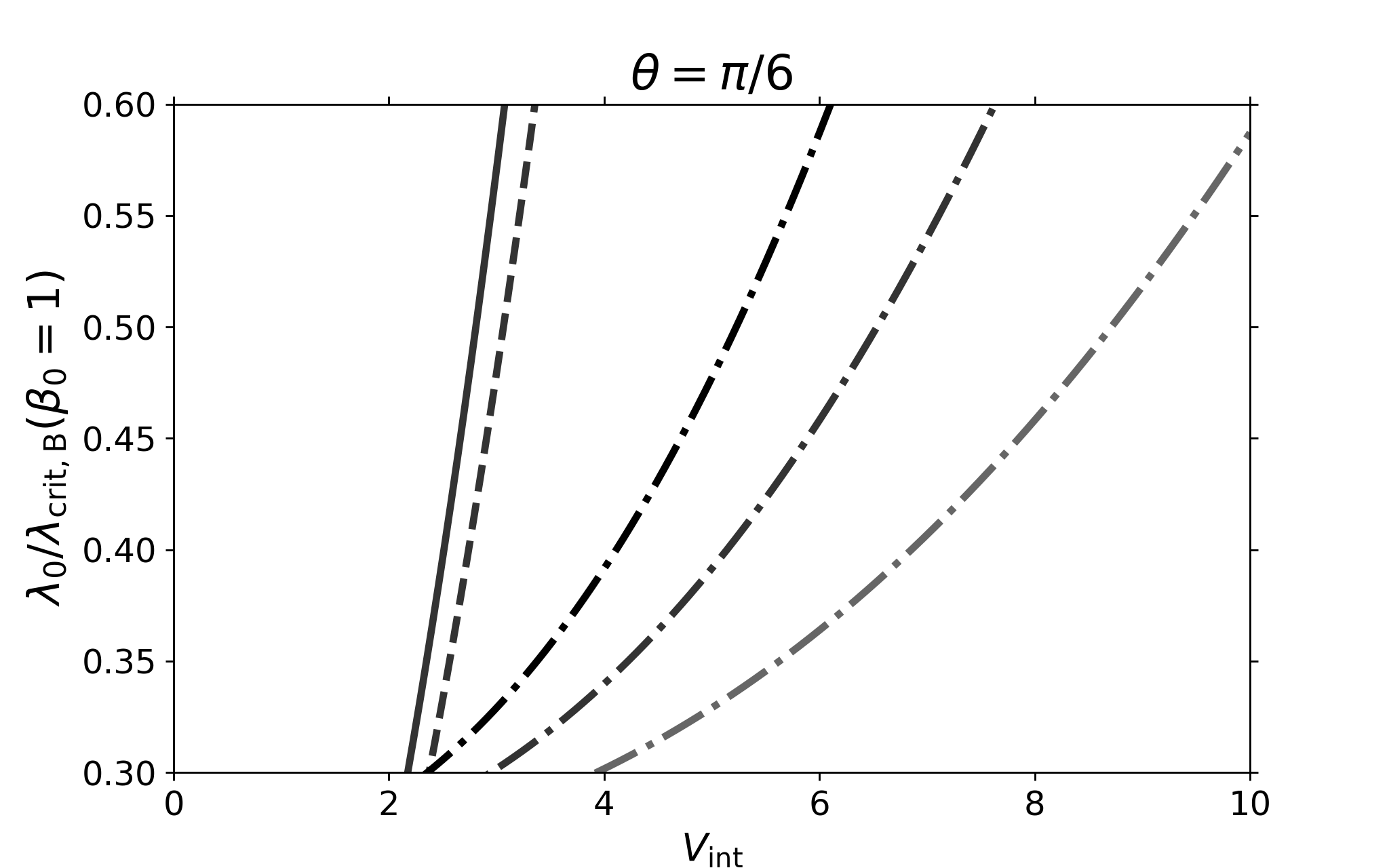}
} \\
\subfigure[]{
\includegraphics[width=0.31\textwidth]{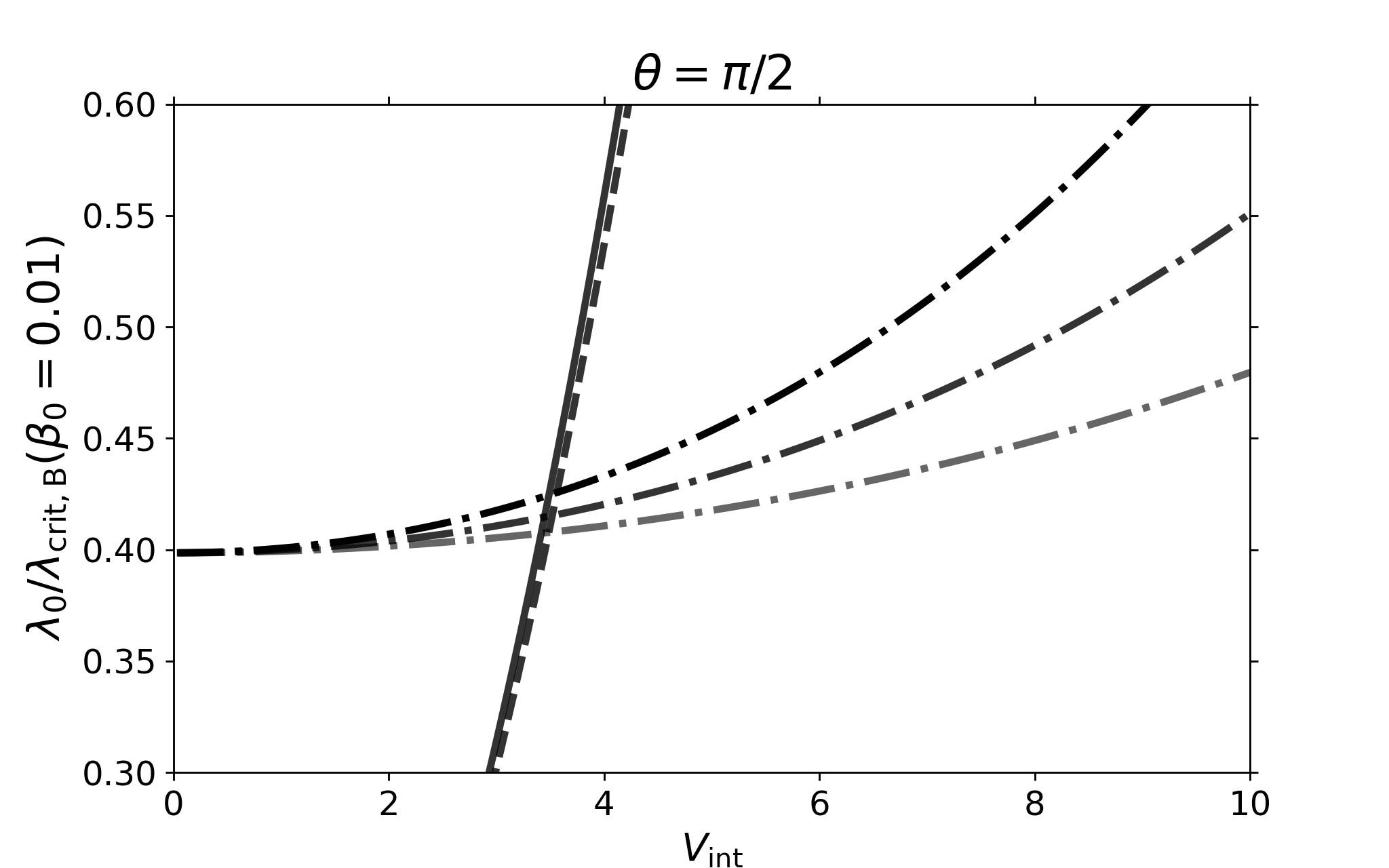}
} &
\subfigure[]{
\includegraphics[width=0.31\textwidth]{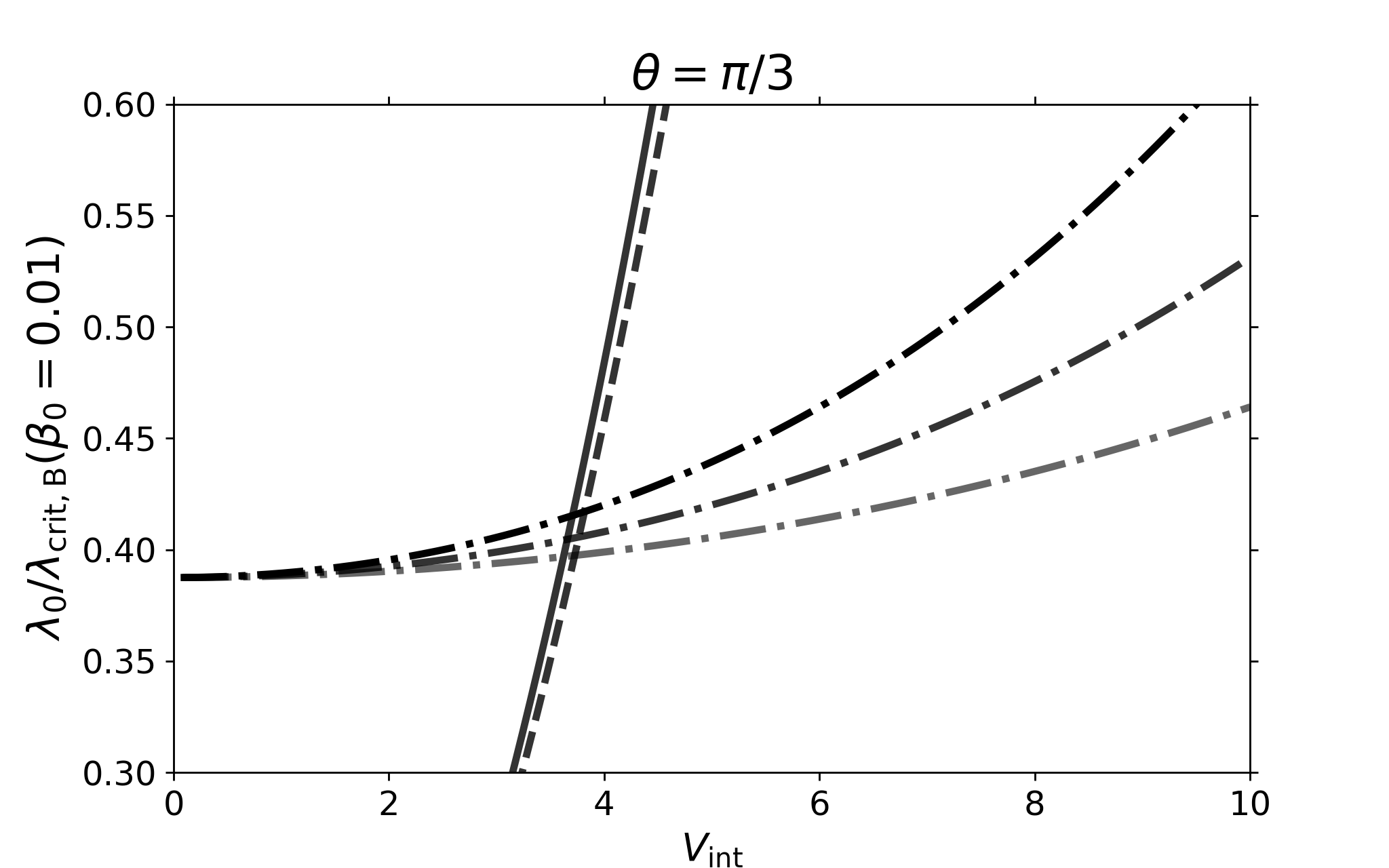}
} &
\subfigure[]{
\includegraphics[width=0.31\textwidth]{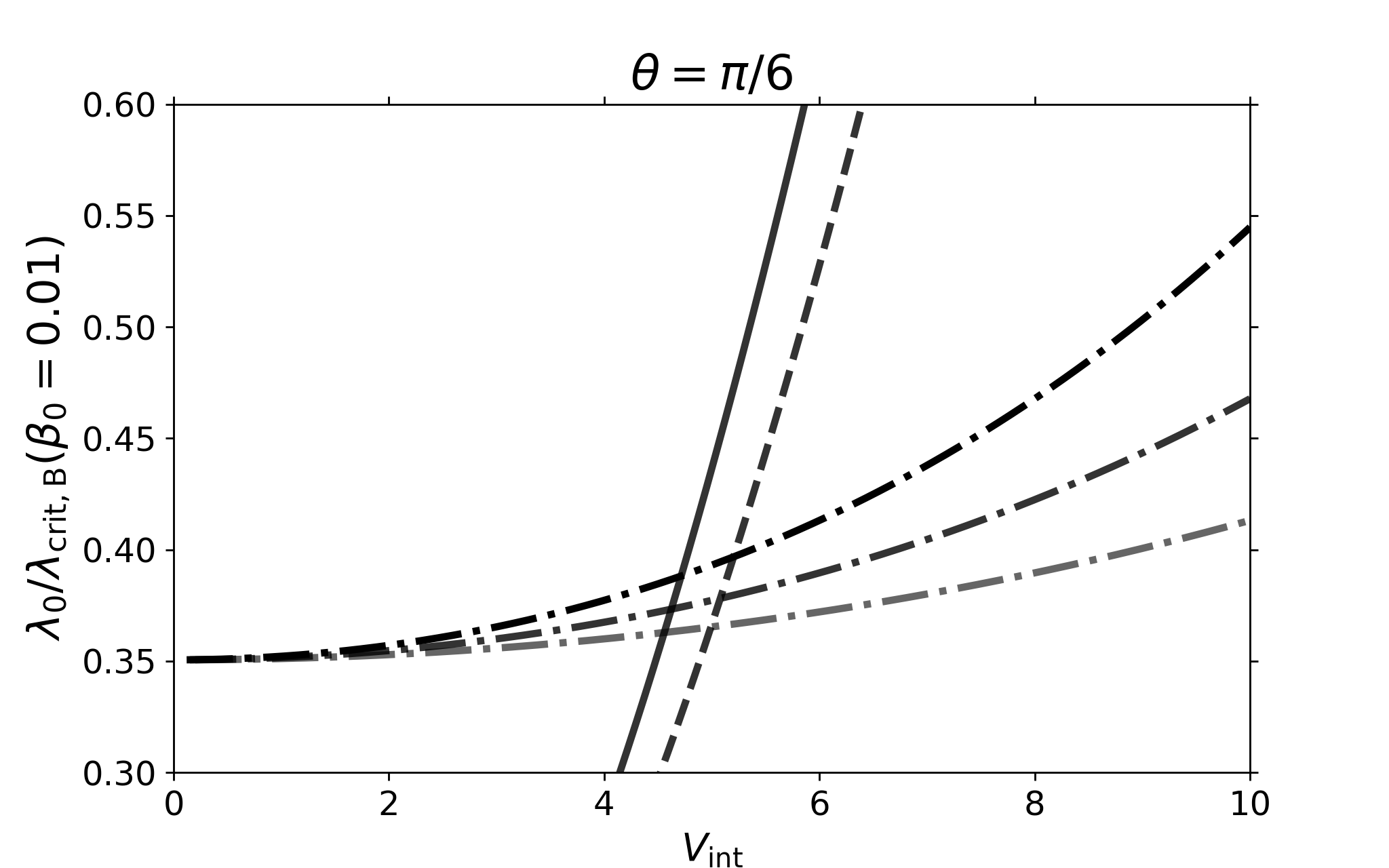}
} 
\end{tabular}
\caption{
Same as figure \ref{fig:stability_for_oblique}, but for the different $\beta_0$. The upper and lower rows correspond to the results for $\beta_0 = 1$ (i.e., $B_0=6\,\mu \mathrm{G}$) and $\beta_0 = 0.01$ (i.e., $B_0=60\,\mu \mathrm{G}$), respectively, while the left-to-right panels show the results for $\theta = \pi/2$, $\theta = \pi/3$, and $\theta = \pi/6$, respectively. For the dash-dotted lines, the geometric factor of the initial filament, $\eta$, is set to its average value for each $\beta_0$: $\eta=0.65$ for $\beta_0=1$ and $\eta=0.30$ for $\beta_0=0.01$. In addition, as in the $\beta_0=0.1$ case, we keep $R_0=0.22\,\mathrm{pc}$ and $\epsilon=0.4$ fixed. 
{Alt text: A set of panels showing the stability outcomes of oblique filament collisions for different initial plasma beta values.}
}
\label{fig:result_of_dif_beta}
\end{figure*}

In the main text, we fixed $\beta_0 = 0.1$ (i.e., $B_0=19\,\mu \mathrm{G}$) and discussed the results under this assumption.
Here, we examine how the collapse–expansion condition (section \ref{sec:condition}) and the escape velocity (section \ref{sec:condition_for_escape_velocity}) depend on $\beta_0$.

Figure \ref{fig:result_of_dif_beta} shows the collapse–expansion boundaries for different values of $\beta_0$. 
To plot the boundaries, we vary the minor-to-major axis ratio of the initial filament, $\eta$, according to the plasma beta, adopting average values of $\eta=0.65$ for $\beta_0=1$ and $\eta=0.30$ for $\beta_0=0.01$. In addition, as in the $\beta_0=0.1$ case, we keep $R_0=0.22\,\mathrm{pc}$ and $\epsilon=0.4$ fixed.
The vertical axis in each panel represents the line mass normalized by the magnetic critical line mass ($\lambda_\mathrm{crit,B}$), which increases with decreasing $\beta_0$ according to equation~(\ref{eq:critical_line_mass}). 
For example, $\lambda_\mathrm{crit,B}$ takes values of $18.7$, $30.6$, and $68.0\,M_\odot\,\mathrm{pc^{-1}}$ for $\beta_0 = 1$, $0.1$, and $0.01$, corresponding to $B_0 = 6$, $19$, and $60\,\mu\mathrm{G}$, respectively \citep{2014ApJ...785...24T}. 
Consequently, filaments with the same criticality ($\lambda_0/\lambda_\mathrm{crit,B}$) correspond to larger line masses under stronger magnetic fields. 
This can be seen by comparing figure~\ref{fig:result_of_dif_beta} (d) with figure~\ref{fig:stability_for_oblique} (a): because $\lambda_\mathrm{crit,B}$ is $68.0\,M_\odot\,\mathrm{pc^{-1}}$ for $\beta=0.01$ and $30.6\,M_\odot\,\mathrm{pc^{-1}}$ for $\beta=0.1$, the line mass in the $\beta=0.01$ models is about 2.2 times larger than in the $\beta=0.1$ models for the same criticality.

When the magnetic field strength increases, the magnetic energy term in equation (\ref{eq:KE_after_PE_ME_GE}) increases accordingly. 
To satisfy the condition $E_\mathrm{tot}=0$, the initial filaments must have a larger line mass. 
As a consequence, only more massive filaments can trigger the collapse mode. 
In fact, in both figure~\ref{fig:result_of_dif_beta} (d) and figure~\ref{fig:stability_for_oblique} (a), the collapse–expansion boundaries lie around a criticality of $\lambda_0/\lambda_{\mathrm{crit,B}}\sim 0.4$ for collision velocities up to a few times $c_s$, whereas the actual line mass is about 2.2 times larger in the $\beta=0.01$ models than in the $\beta=0.1$ models.
In addition, as shown in figures \ref{fig:result_of_dif_beta} (d)-(f), the collapse–expansion boundaries shift downward as $\theta$ decreases. 
This is because a smaller $\theta$ leads to a larger mass of the compressed cloud, as discussed in section \ref{sec:inclination_angle_dependence}, which increases the gravitational energy.

Conversely, when the magnetic field is weaker, comparison of figure~\ref{fig:result_of_dif_beta} (a) with figure~\ref{fig:stability_for_oblique} (a) shows that the line mass required to satisfy $E_\mathrm{tot}=0$ becomes smaller, making even low-line-mass filaments more prone to collapse.
In addition, in figure \ref{fig:result_of_dif_beta} (a)-(c), the trend that the collapse–expansion boundaries shift downward as $\theta$ decreases remains unchanged.

Figure \ref{fig:result_of_dif_beta} also shows the escape velocities for different $\beta_0$, indicated by the solid lines for each collision angle.
The escape velocity range shifts toward higher values as $\beta_0$ decreases, although the exact range depends on the collision angle.
This arises because, when comparing models with the same criticality at different values of $\beta_0$, a smaller $\beta_0$ corresponds to more massive filaments, which in turn leads to a larger total mass of the compressed cloud.
Nevertheless, within the range of $\beta_0 = 1$–$0.01$ and collision angles from $\theta = \pi/2$ to $\theta = \pi/6$, the escape velocities remain within $V_\mathrm{esc} \simeq 1.5\,c_s$–$6\,c_s$, indicating that the non-colliding segments can easily escape from the gravitational potential of the compressed cloud.

\section{Derivation for the accurate escape velocity}\label{sec:Center_of_mass_of_trapezoid}

\begin{figure}
\begin{center}

\includegraphics[keepaspectratio,scale=0.4]{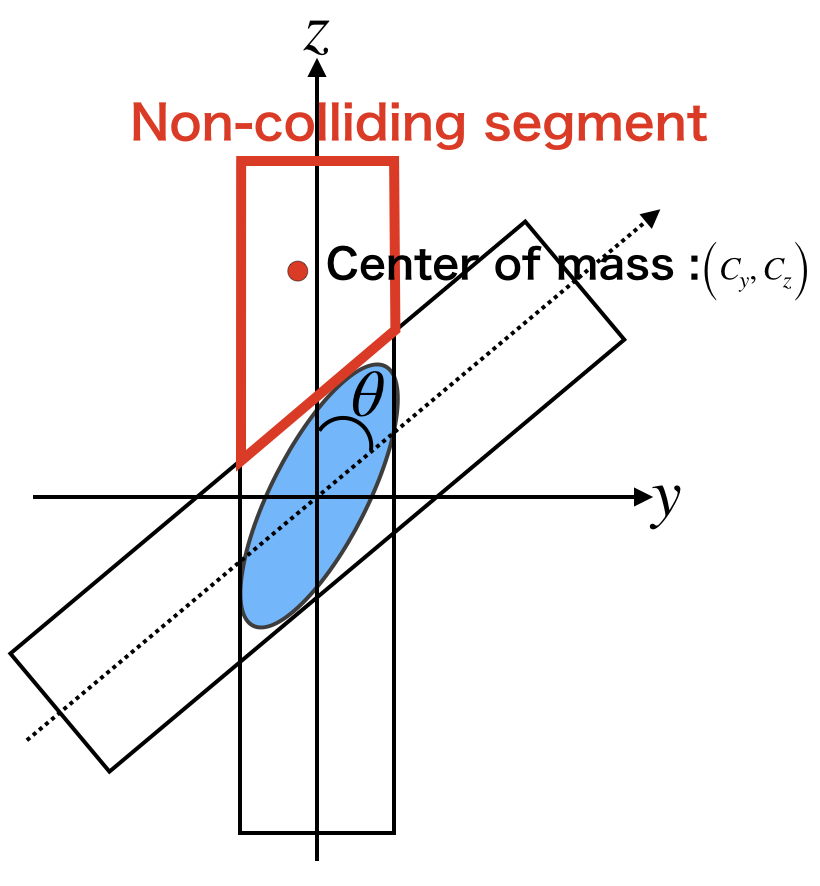}
\end{center}
      \caption{
      Schematic view of the non-colliding segment on the $x = 0$ plane. The non-colliding segments move forward or backward along the line of sight at a velocity of $V_\mathrm{int}$. One of them is illustrated as the red trapezoid, whose center of mass is located at $(C_y\, ,C_z)$. 
      {Alt text: A schematic diagram on the x=0 plane showing the non-colliding filament segments.}
      }
    \label{fig:Non_colliding_segment}
        
\end{figure}

In this appendix, we derive a more accurate expression for the escape velocity by evaluating the gravitational potential of the compressed cloud, taking into account its dependence on the collision angle, and by estimating the center of mass of the non-colliding segment, which is modeled as a trapezoid.
Using the distance between the center of mass of the trapezoidal segment and the origin, which coincides with the center of mass of the compressed cloud, we evaluate the gravitational potential of the triaxial ellipsoid and thereby estimate the escape velocity.

The compressed cloud is modeled as the Jacobi ellipsoid \citep[see Eq(70) in][]{1969efe..book.....C}; thus, the gravitational potential at $(x,\,y,\,z)$ outside this cloud is expressed as
\begin{equation}
    \displaystyle \phi = \pi G\rho a_1a_2 a_3\int^{\infty}_{k}\frac{du}{\Delta}\left[1- \left(\frac{x^2}{a^2_1+u}+\frac{y^2}{a^2_2+u}+\frac{z^2}{a^2_3+u}\right)\right],
\end{equation}
where $\Delta\equiv \left[ \left(a^2_1+u \right) \left(a^2_2+u \right) \left(a^2_3+u \right)  \right]^{1/2}$ and $k$ is obtained from
\begin{equation}
    \frac{x^2}{a^2_1+k}+\frac{y^2}{a^2_2+k}+\frac{z^2}{a^2_3+k}=1.
\end{equation}

In this analysis, we assume that the non-colliding segment has a trapezoidal prism shape and a uniform density.
In addition to this, the escape velocity of the non-colliding segment is approximated by that of the gas at the center of mass of the trapezoid, as shown figure \ref{fig:Non_colliding_segment}.
To derive the location of the center of mass of the trapezoidal prism, we use the centroid formula for a polygon.
In figure \ref{fig:Non_colliding_segment}, we focus on one of the four non-colliding segments, specifically the one located in the $z > 0$ region of the filament whose long axis is fixed along the $z$-axis.
In addition to this, the trapezoid is assumed to represent the projected shape of the non-colliding segment on the $y-z$ plane, and the vertex order is taken in a counterclockwise manner.
Therefore, the center of mass of this non-colliding segment is represented as $(0,\,C_y,\,C_z)$. 
The vertices of the trapezoid are defined as  
$(y_1,z_1)\equiv (-R_0, L_\mathrm{fil}/2)$,  
$(y_2,z_2)\equiv (R_0, L_\mathrm{fil}/2)$,  
$(y_3,z_3)\equiv (R_0, R_0/\tan\theta + R_0/\sin\theta)$, and  
$(y_4,z_4)\equiv (-R_0, -R_0/\tan\theta + R_0/\sin\theta)$.
Then, the centroid of the trapezoid is calculated using the centroid formula for a polygon \citep[see][]{1988_bourke_centroid}:
\begin{equation}
    C_y=\frac{1}{6A}\sum^{4}_{i=1}(y_i+y_{i+1})(y_iz_{i+1}-y_{i+1}z_i),
\end{equation}
\begin{equation}
    C_z=\frac{1}{6A}\sum^{4}_{i=1}(z_i+z_{i+1})(y_iz_{i+1}-y_{i+1}z_i),
\end{equation}
where $A$ is given by
\begin{equation}
    A=\frac{1}{2}\sum^{4}_{i=1}(y_iz_{i+1}-y_{i+1}z_i),
\end{equation}
and for $i=4$, the index is taken cyclically as $(y_{i+1},z_{i+1})=(y_1,z_1)$.
It should be noted that this formula is valid as long as $R_0\left({\cos\theta+1}\right)/{\sin\theta}\le {L_\mathrm{fil}}/{2}$; otherwise, the non-colliding segment no longer forms a trapezoid.
Then, the centroid $(C_y, C_z)$ of the trapezoid can be determined, and the angle between the centroid position and the $z$-axis is given by
\begin{equation}
\xi = \arctan\left(\frac{|C_y|}{C_z}\right).
\end{equation}

In the coordinate system aligned with the principal axes of the Jacobi ellipsoid, i.e., $a_1 \parallel x_1$, $a_2 \parallel x_2$, and $a_3 \parallel x_3$, the position of the centroid is given by
\begin{equation}
(x_1, x_2, x_3) = \left( r \cos(\xi + \theta/2),\ r \sin(\xi + \theta/2),\ 0 \right),
\end{equation}
where $r=\sqrt{C^2_y+C^2_z}$.

Finally, the gravitational potential of the Jacobi spheroid is given, using the relation $M={(4\pi/3)} \rho a_1a_2a_3\simeq 4\lambda_0R_0/\sin\theta$ as  
\begin{equation}\label{eq:gravitational_potential}
\begin{split}
    \phi &= \frac{3G\lambda_0R_0}{\sin\theta}\int^{\infty}_{k}\frac{du}{\Delta}\left(1- \sum^{3}_{i=1}\frac{x^2_i}{a^2_i+u}\right),
    %\phi' &= \frac{3}{2\pi}\frac{\lambda'_0}{\sin\theta}\int^{\infty}_{k'}\frac{du'}{\Delta'}\left[1- \left(\frac{(r'\cos(\xi+\theta/2))^2}{a'^2_1+u'}+\frac{(r'\sin(\xi+\theta/2))^2}{a'^2_2+u'}+\frac{0^2}{a'^2_3+u'}\right)\right],
\end{split}
\end{equation}
where $k$ is determined by solving
\begin{equation}\label{eq:sufficient_condition}
    \frac{(r\cos(\xi+\theta/2))^2}{a^2_1+k}+\frac{(r\sin(\xi+\theta/2))^2}{a^2_2+k}
    %+\frac{0^2}{a'^2_3+k'}
=1.
\end{equation}
Therefore, the escape velocity, defined as $V_\mathrm{esc} = \sqrt{-2\phi}$, is obtained by solving equations (\ref{eq:gravitational_potential}) and (\ref{eq:sufficient_condition}).

In figures~\ref{fig:stability_for_oblique} and \ref{fig:result_of_dif_beta}, the more accurate escape velocities are shown as dashed lines.
As the collision angle decreases, the deviation from the solid lines becomes larger due to the shape of the shocked cloud departing from a sphere.
However, these deviations are not significant; therefore, we adopt the simplified escape velocity in the main text, as given by equation~(\ref{eq:escape_vel_hub}).

%\section{Why we can not observe edge collapse without parallel collisions} 

% Any journal's BST file (e.g., apj.bst) can be used as PASJ's BST is unavailable.    
% \bibliographystyle{****}
% \bibliography{****}
%\bibliography{accretion_ffc}{}
%\bibliographystyle{aasjournal}
%\bibliographystyle{}
%\begin{thebibliography}{}
% Journals(e.g. A\&A,ApJ,AJ,NMRAS,PASP ...)
% Authors, Year, Journal, Vol#, Page# 
%\bibitem[Aauthor et al.(2001)]{key-1}
%  Aauthor, A., Bauthor, B., \& Cauthor, C.\ 2001, PASJ, 53, 000 
%\bibitem[Aauthor \& Bauthor(2003a)]{key-2}
%  Aauthor, A., \& Bauthor, B.\ 2003a, PASJ, 55, 000 
%\bibitem[Aauthor \& Bauthor(2003b)]{key-3}
%  Aauthor, A., \& Bauthor, B.\ 2003b, PASJ, 55, 000 
%\bibitem[Aauthor, Cauthor, and Dauthor(2000)]{key-3}
%  Aauthor, A., Cauthor, C., \& Dauthor, D.\ 2000, PASJ, vol, page   
% Books
%\bibitem[Aauthor \& Eauthor(2003b)]{key-4}
%  Aauthor, A., \& Euthor, E.\ 2003b, Name of Book (Tokyo: Publisher) ch.0    
% Editorial Books
%\bibitem[Dauthor(2001)]{key-5}
%  Dauthor A.~A.\ 2001, in Name of Book, ed.\  D.~Editor (Tokyo: Publisher), 00 
%\end{thebibliography}

\end{document}